\documentclass{aa}  
\usepackage[pdfencoding=auto,psdextra,hidelinks]{hyperref}
\newcommand{\exo}[3]{{\em #1#2#3}}
\newcommand{\citaeg}[2]{(#1\ \citealt{#2})}
\usepackage{graphicx}
\usepackage{svg}
\usepackage{txfonts}
\usepackage{lipsum}
\usepackage{subcaption}         
\usepackage{lscape}            
\usepackage{placeins}           

\begin{document}

   \title{Efficient reduction of stellar contamination and noise in planetary transmission spectra using neural networks}
   \titlerunning{Reducing stellar contamination with DAEs}

   \author{David S. Duque-Castaño\thanks{Corresponding author: \href{mailto:dsantiago.duque@udea.edu.co}{dsantiago.duque@udea.edu.co}}
        \and Lauren Flor-Torres
        \and Jorge I. Zuluaga}

   \institute{SEAP/FACom, Instituto de F\'{\i}sica - FCEN, Universidad de Antioquia, Calle 70 No. 52-21, Medell\'in, Colombia\\
   }

   \date{Received January XX, 2026; In original form January XX, 2026}
 
  \abstract
      {The characterization of exoplanetary atmospheres has been transformed by the James Webb Space Telescope (JWST), whose infrared sensitivity enables transmission spectroscopy at unprecedented precision. However, stellar heterogeneities (e.g., spots and faculae) remain a dominant source of contamination that can bias atmospheric retrievals if not properly corrected.}%
      {We present a methodology for reducing stellar contamination and instrument-specific noise from exoplanet transmission spectra using neural networks, in particular the so-called denoising autoencoders (DAEs). Our goals are to enable fast, accurate corrections that improve the reliability of atmospheric parameter retrievals and to promote the use of unsupervised algorithms for efficient data processing.}%
      {We designed and trained DAE architectures using large synthetic datasets of terrestrial (\exo{TRAPPIST-1}{}{e} analogs) and sub-Neptune (\exo{K2}{-18}{b} analogs) planets.
      Atmospheric retrieval experiments were then performed on contaminated spectra in order to compare our deep-learning approach against standard correction methods in terms of accuracy and computational cost.}%
      {Our autoencoders successfully reconstruct uncontaminated spectra, preserving essential molecular features even in low signal-to-noise ratio regimes. In retrieval tests, the DAE preprocessing yields atmospheric parameter estimates broadly comparable to those obtained with simultaneous stellar-contamination fitting. Notably, our method maintains a much lower computational cost, approximately one order of magnitude smaller.}%
      {These results demonstrate that DAEs outperform conventional correction methods in computational efficiency while maintaining high accuracy, paving the way for their integration into future atmospheric characterization pipelines for both rocky and sub-Neptune exoplanets.}


   \maketitle
\nolinenumbers

\section{Introduction}
\label{sec:introduction}

The characterization of exoplanetary atmospheres began nearly two decades ago with the pioneering detection of sodium in the atmosphere of the hot Jupiter \exo{HD}{ 209458}{b} \citep{charbonneauDetectionExtrasolarPlanet2002}. Since then, successive observations with Hubble, Spitzer, and ground-based telescopes have revealed signatures of various compounds in the atmospheres of giant exoplanets, including water vapor \citep{tinettiWaterVapourAtmosphere2007} and carbon monoxide \citep{snellenOrbitalMotionAbsolute2010}. In recent times, the James Webb Space Telescope (JWST) has greatly improved our capabilities for transmission spectroscopy, allowing us to detect CO$_2$, H$_2$O, Na, and CO within even individual transit spectra \citaeg{see e.g.,}{rustamkulovEarlyReleaseScience2023}. 

Beyond gas giants, JWST has also begun probing a broader diversity of planets, from rocky worlds to sub-Neptunes orbiting low-mass stars. These systems span a wide range of atmospheric regimes and observational conditions, making them valuable test cases for transmission spectroscopy. 

Recently, tentative evidence for compounds such as dimethyl sulfide (DMS), a potential biosignature \citep{madhusudhanCarbonbearingMoleculesPossible2023, madhusudhanNewConstraintsDMS2025}, has been reported. Transmission spectra showing signatures of CH$_4$, CO$_2$, and H$_2$O have also been reported for Earth-sized to sub-Neptune-sized planets \citep{bennekeJWSTRevealsCH$_4$2024}, and a recent study of \exo{LHS}{ 1140}{b} using JWST/NIRISS points toward an atmosphere dominated by N$_2$, with evidence of Rayleigh scattering and the exclusion of a H$_2$-rich atmosphere \citep{cadieuxTransmissionSpectroscopyHabitable2024}.

One of the most significant obstacles in analyzing planetary transmission spectra is the so-called transit light source effect (TLS), also known as stellar contamination.\footnote{Hereafter, and for the sake of clarity, we use the acronym TLS and stellar contamination interchangeably to refer to the same effect.} This effect is produced by the unresolved presence of photospheric heterogeneities such as cool spots and hot faculae
\citep{rackhamTransitLightSource2018,rackhamTransitLightSource2019,rackhamEffectStellarContamination2023}. This is particularly important when analyzing the transmission spectra of rocky exoplanets orbiting M dwarfs, which exhibit strong magnetic activity \citaeg{see e.g.,}{iyerInfluenceStellarContamination2020}. Stellar contamination was clearly corroborated by the JWST/NIRISS observations of \exo{TRAPPIST-1}{}{b} transits \citep{limAtmosphericReconnaissanceTRAPPIST12023}. Moreover, unocculted faculae have been independently identified, using FORS2/VLT, on WASP-69 during the transit of one of its planets, \exo{WASP-69}{}{b} \citep{petitditdelarocheDetectionFaculaeTransit2024}. 

Several approaches have been proposed to mitigate TLS. Thus, for instance, in multi-planet systems, a model-independent method that uses consecutive transits of sibling planets has achieved up to a 2.5-fold reduction in contamination of the \exo{TRAPPIST-1}{}{c} transmission spectrum \citep{rathckeStellarContaminationCorrection2025}. In other scenarios, TLS can be incorporated directly into atmospheric retrieval frameworks \citaeg{see e.g.,}{macdonaldPOSEIDONMultidimensionalAtmospheric2023}. Other studies have relied on out-of-transit stellar spectra to characterize surface heterogeneities and attempt to remove their contaminating influence on transmission spectra (see \citealt{rackhamRobustCorrectionsStellar2024}, hereafter RW2024, and \citealt{cadieuxTransmissionSpectroscopyHabitable2024}). 

Despite these efforts, several key challenges remain. 
First, current stellar spectral models frequently struggle to accurately represent low-mass stars' spectra (see e.g., RW2024).
Second, the reduction of stellar contamination in retrieval procedures is too model-dependent.
Lastly, a significant overlap between stellar and planetary parameters complicates signal separation \citaeg{see e.g.,}{iyerInfluenceStellarContamination2020}. Therefore, with traditional methodologies, improved stellar modeling and further observations are essential for a thorough understanding of these heterogeneities and their effects on transmission spectra.

In this work, we address the aforementioned challenges by proposing a novel neural-network-based methodology. For this purpose, we designed several specialized neural networks, specifically the so-called denoising autoencoders (DAEs), that capture the essential structure of simulated transmission spectra and can reconstruct the intrinsic planetary signal from noisy, contaminated transmission spectra before applying a standard retrieval algorithm. 

It is important to stress that we are not proposing to bypass Bayesian retrieval, to emulate posterior distributions, or to construct a surrogate forward model. Instead, our DAE strategy is developed as a critical preprocessing stage before the application of other tools designed for retrieval purposes. For a short review of the more general application of machine learning algorithms, and in particular of autoencoders, in transmission spectroscopy and other areas of signal processing in astronomy, see \autoref{app:ML}.

This work is organized as follows. In \autoref{sec:stellar_contamination}, we summarize the TLS formalism and introduce the wavelength-dependent contamination factor, $\epsilon_\lambda$, used throughout this work to generate controlled contaminated spectra. In \autoref{sec:G-DAE} and \autoref{subsec:gdae_evaluation}, we present our first numerical experiment: the design, training, and evaluation of a general stellar-contamination denoising autoencoder tailored to terrestrial-planet transmission spectra, including tests with realistic JWST/NIRSpec PRISM uncertainties and controlled retrieval comparisons with \textsc{POSEIDON}. In \autoref{sec:other_planets}, we extend the methodology to sub-Neptune transmission spectra and assess performance across contamination and noise regimes. We discuss limitations, domain generalization, and practical considerations for real observations in \autoref{sec:discussion}. Finally, we summarize the main findings and outline future directions in \autoref{sec:summary_and_conclusions}.

\section{Stellar contamination}
\label{sec:stellar_contamination}

Stellar contamination, or the TLS effect, is modeled as a wavelength-dependent modification of the observed planetary radius as

\begin{equation}
\left[\frac{R_p(\lambda)}{R_\star}\right]^2_\mathrm{obs} = \left[\frac{R_p(\lambda)}{R_\star}\right]^2 \epsilon_\lambda,
\end{equation}
where $\epsilon_\lambda$ is a multiplicative contamination factor that transforms an uncontaminated transmission spectrum ($[R_p/R_\star]$ on the right-hand side) into a contaminated one (on the left-hand side). 

Under simplified assumptions (for a complete derivation see \autoref{app:stellar_contamination}), $\epsilon_\lambda$ can be expressed in terms of the area covered by spots, $f_\mathrm{spot}$, or faculae, $f_\mathrm{fac}$, and the fraction of the transit chord covered by those heterogeneities, $c_\mathrm{spot}$ and $c_\mathrm{fac}$, using the expression

\begin{equation}
\epsilon_{\lambda} = \frac{(1 - c_{\mathrm{spot}} - c_{\mathrm{fac}}) F_\mathrm{phot}(\lambda) + c_{\mathrm{spot}} F_\mathrm{spot}(\lambda) + c_{\mathrm{fac}} F_\mathrm{fac}(\lambda)}{(1 - f_{\mathrm{spot}} - f_{\mathrm{fac}}) F_\mathrm{phot}(\lambda) + f_{\mathrm{spot}} F_\mathrm{spot}(\lambda) + f_{\mathrm{fac}} F_\mathrm{fac}(\lambda)}.\label{eq:epsilon}
\end{equation} Here $F_\mathrm{phot}(\lambda)$, $F_\mathrm{spot}(\lambda)$, and $F_\mathrm{fac}(\lambda)$ are the outgoing fluxes from the photosphere, the spot, and the facula regions, respectively.  

The goal of several mitigation strategies has been to estimate the values of the unknown parameters $f_\mathrm{spot}$, $f_\mathrm{fac}$, $c_\mathrm{spot}$, and $c_\mathrm{fac}$ in order to better reconstruct a signal compatible with a given chemistry (RW2024, \citealt{cadieuxTransmissionSpectroscopyHabitable2024}). Despite these advances, significant limitations persist, including the difficulty that stellar models have in accurately reproducing M-dwarf spectra and the substantial degeneracy between stellar and planetary parameters \citep{iyerInfluenceStellarContamination2020}.

\bigskip

In this study, we used this formula not to correct a given observed spectrum but to generate thousands of contaminated transmission spectra that serve as training data for a custom neural network, specifically a DAE. The general denoising autoencoder (G-DAE, pronounced ``yee-dai'') constructed here is able to capture the key characteristics of the original spectrum while mitigating stellar contamination. In the sections that follow, we present the architecture of our G-DAE and evaluate its performance on realistically simulated data.

\section{A general stellar contamination DAE}
\label{sec:G-DAE}

For our first numerical experiment, we explored the case of Earth-like exoplanets orbiting M dwarfs and containing potential biosignatures and/or bioindicators, as we did in \cite{duque-castanoMachineassistedClassificationPotential2025}. For this purpose, we modeled the case of \exo{TRAPPIST-1}{}{e} for which we assumed a diverse set of atmospheric compositions having at least five molecular species, CO$_2$, N$_2$ (fill gases), CH$_4$, O$_3$, and H$_2$O, which produce the strongest spectral signatures \citaeg{see e.g.,}{kalteneggerTRANSITSEARTHLIKEPLANETS2009, lustig-yaegerEarthTransitingExoplanet2023}. For illustration, in \autoref{app:uncontaminated_spectra} we present a representative theoretical spectrum of such a planet (see \autoref{fig:trappist1e_analogue_spectrum}), which we used to train our models.

\subsection{Model architecture}
\label{subsec:gdae_architecture}

Our first G-DAE was developed in \textsc{Keras} \citep{chollet2015keras} with a \textsc{TensorFlow} backend \citep{tensorflow2015-whitepaper}.  The network was trained entirely in this transformed space.  During inference, the network receives only the transformed noisy and contaminated spectrum (see below \autoref{subsec:spectra_normalization}). The clean spectrum was used as a reconstruction target during training and as a reference for post hoc evaluation, but it is not supplied to the trained network during inference.

The network processes the data through a symmetric encoder--decoder: the encoder comprises three fully connected layers of 512 units, followed by two layers of 300 units; the decoder mirrors this architecture and ends with a linear layer that projects back to 385 outputs. All hidden units use the Swish activation function $f(x)=x/(1+e^{-x})$ \citaeg{see e.g.,}{geronHandsonMachineLearning2023}. To mitigate overfitting, we applied dropout with a rate of 0.5 after every hidden layer and L2 (weight-decay) regularization with a coefficient $\lambda=10^{-7}$ on all dense kernels except the final output layer. \autoref{sec:autoencoders} provides a brief primer on autoencoders and their application in astronomy, along with a discussion on the main noise sources affecting planetary transmission spectroscopy. There, we provide, specifically in \autoref{fig:autoencoder_general}, a schematic representation of the G-DAE network architecture used in this work.

The G-DAE hyperparameters were selected through a grid search over network depth and width, dropout rate, and the L2 coefficient, training on the full spectra set (see below) and selecting the configuration that minimized the ``mean absolute error'' (MAE). As noted in \autoref{sec:autoencoders}, the MAE is computed between the reconstructed spectrum and its corresponding clean target before adding stellar contamination and noise; the L2 penalty is added through the layer regularizer while MAE remains the primary reconstruction loss.

\subsection{Spectra normalization}
\label{subsec:spectra_normalization}

All spectra (clean, contaminated, and noisy), denoted by $D(\lambda)$, were first normalized by a reference transit depth, $D_{\rm ref}$, such that $d_{\rm obs}(\lambda) \equiv D(\lambda)/D_{\rm ref}$. This reference depth was tied to a fixed planetary radius (for instance, that of \exo{TRAPPIST-1}{}{e}). Stellar contamination appears as a multiplicative factor, $D(\lambda)_{cont}=D(\lambda)\epsilon_\lambda$, so taking the logarithm of the normalized signal disentangles the planetary spectrum from the contamination (logarithmic formulations are frequently employed to handle multiplicative spectral contributions \citaeg{see e.g.,}{kjaersgaardTAUNeuralNetwork2023, espinozaJWSTTSTDREAMSNIRSpec2025a}), yielding
$$
X_{\rm obs}(\lambda) = \ln d_\mathrm{obs} =
\ln\left[\frac{D(\lambda)}{D_{\rm ref}}\right] + \ln \epsilon_\lambda.
$$

For the algorithmic implementation, $X$ denotes the logarithm of the corresponding normalized spectrum and is the exact quantity provided to, and then returned by, the G-DAE. This formulation enables the G-DAE to learn spectral features that do not depend on the planetary radius. In particular, the network is neither tied to an absolute scale (which would reduce its ability to generalize to other spectra) nor forced into an arbitrary interval (such as 0–1), which would remove information about abundances. After reconstruction, the G-DAE produces a vector $\hat{X}$, from which we recover the signal in physical units via
$$
\widehat{D}(\lambda) = D_{\rm ref}\exp\left[\widehat{X}(\lambda)\right].
$$

During G-DAE training, the same planetary radius was always assumed. For real data, however, $D_{\rm ref}$ must be replaced in both the normalization and the back-conversion by $D_p$, the measured transit depth of the planet.

\subsection{Model training and validation}
\label{subsec:gdae_training}

To generate the training set, we used the same methodology as described in \citet{duque-castanoMachineassistedClassificationPotential2025}. For completeness, we briefly outline the generation methodology below.

In our training set, atmospheric compositions contained none, one, two, or all of the key molecules, CH$_4$, O$_3$, and H$_2$O. Mixing ratios for each species were drawn from eight discrete values distributed log-uniformly between \(10^{-8}\) and \(10^{-1}\). These were combined with three representative isothermal atmospheric temperatures of 200~K, 287~K, and 400~K. The baseline fill gas atmosphere consisted of N$_2$ plus CO$_2$ with a mixing ratio of \(10^{-3}\), representing an intermediate case between modern Earth and early Earth conditions \citep{kalteneggerSpectralEvolutionEarthlike2007}. We also included an airless planet that produces flat spectra with no atmospheric absorption. All atmospheres were assumed to have an Earth-like surface pressure of 1 bar, and their vertical structure was represented with a 100-layer model.

For each base spectrum (fixed chemistry, surface pressure, and temperature profile), multiple noisy realizations were produced at five signal-to-noise ratio (S/N) values: 0 (noise-free), 1, 3, 6, and 10. Gaussian noise with wavelength-independent standard deviation $\sigma_{0}$ was added to all spectral points, where $\sigma_0 = S_0/(S/N)$, with $S_0$ the amplitude of the strongest CO$_2$ absorption feature in the uncontaminated spectrum over the spectral range considered. Stellar contamination was simulated by multiplying each spectrum by precomputed wavelength-dependent $\epsilon_\lambda$ (see \autoref{eq:epsilon}). For this purpose, we generated two parallel families of contamination curves for \exo{TRAPPIST-1}{}{}, one based on PHOENIX \citep{ husserNewExtensiveLibrary2013} within the \textsc{POSEIDON} framework \citaeg{see e.g.,}{macdonaldPOSEIDONMultidimensionalAtmospheric2023} and another based on SPHINX \citep{iyerSPHINXMdwarfSpectral2023} by directly evaluating \autoref{eq:epsilon} after interpolating the stellar spectra in effective temperature and regridding all components onto the common wavelength mesh. In all cases, we assumed, for simplicity, that the transit chord contains no heterogeneities, i.e., $c_\mathrm{spot}=c_\mathrm{fac}=0$. For the coverage fractions, we combined three values of $f_\mathrm{spot}=0.01$, 0.08, and 0.26, and three values of $f_\mathrm{fac}=0.08$, 0.54, and 0.70, following the suggestions by \citet{rackhamTransitLightSource2018}.

In the terrestrial dataset, each base spectrum was combined with 19 stellar cases: nine contaminated configurations generated with PHOENIX, nine generated with SPHINX, and one uncontaminated case. For example, a family of spectra with one molecular species at S/N = 1 was constructed from eight possible mixing ratios for that molecule, combined with three different temperatures, giving $8\times3 = 24$ base spectra. The same configuration was then replicated across 19 stellar cases and 200 noisy realizations, yielding $[(8\times3)\times19]\times200 = 91{,}200$ spectra. A complete breakdown of the molecular configurations, noise levels, and airless scenarios used to construct the dataset, with counts accounting for all stellar contamination cases and noisy realizations, is provided in \autoref{tab:Gen_DAE_dataset}. Altogether, the combination of all variables and their effects resulted in a training dataset containing $3{,}605{,}592$ spectra.

To fit the model, the data were split into 80\% for training and 20\% for validation. Learning was carried out using the Adam optimizer \citaeg{see e.g.,}{geronHandsonMachineLearning2023} with a learning rate of $10^{-5}$, a batch size of 64, MAE loss, early stopping with a patience of five epochs, and a maximum of 100 epochs. Most of these optimization parameters were determined by systematic experimentation.

\begin{figure*}
    \centering
    \includegraphics[width=1.0\linewidth
    ]{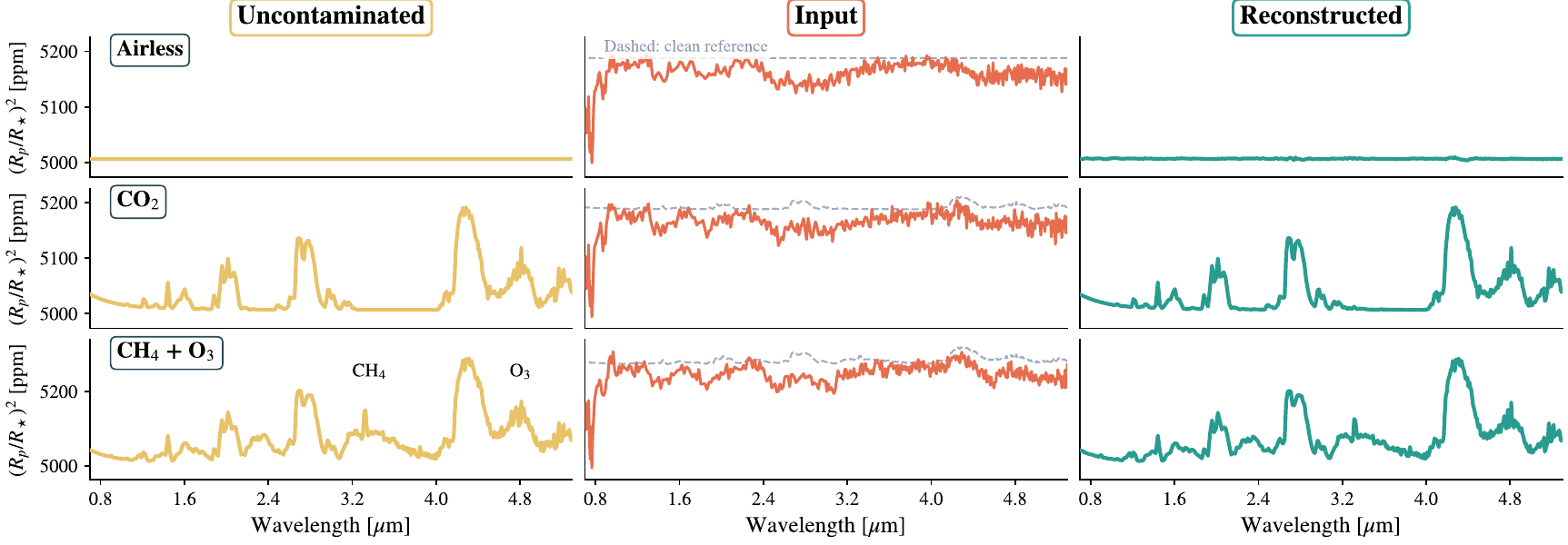}
    \caption{Performance of the general autoencoder (G-DAE) for terrestrial-like planets in three atmospheric cases: airless (first row), single-molecule (second row), and complex atmospheres with fill gases and two bioindicators (third row). The input contaminated spectra have S/N = 3 (see text for details), with contamination levels $f_{\mathrm{spot}} = 0.08$ and $f_{\mathrm{fac}} = 0.54$.}
    \label{fig:spectrum_cleaning}
\end{figure*}

After training, we inspected the G-DAE performance using reserved datasets. First, we verified that the model achieved satisfactory reconstruction metrics during validation and preserved the main spectral features. For the latter purpose, we visually inspected random reconstructed spectra. In \autoref{fig:spectrum_cleaning}, we show three characteristic examples of spectra having none, one, and two molecular species. The results indicate strong reconstruction performance, at least at a visual level. Despite the significant contamination (compare the original spectra in the first column with the contaminated one in the second column), the G-DAE can reconstruct the original signal with high fidelity.

However, visual inspection alone is not sufficient to judge the performance of the G-DAE. As a first quantitative assessment, we report two standard measures of reconstruction accuracy: the signal-normalized mean squared error, defined as ${\rm nMSE}=100\,{\rm MSE}/\Delta D_{\rm ref}^{2}$, and the coefficient of determination, $R^2$. Here, $\Delta D_{\rm ref}$ is the peak-to-baseline amplitude of the uncontaminated reference signal used in the corresponding synthetic model. Although these metrics directly quantify the agreement between the reconstructed and clean spectra, they do not account for the predictive uncertainties associated with the reconstruction. A more complete assessment incorporating these uncertainties is therefore presented in the model evaluation  (\autoref{subsec:gdae_evaluation}).

\renewcommand{\thefootnote}{\alph{footnote}}
The network achieved ${\rm nMSE}=0.448\%$\footnote{Although our loss function during the learning process was the MAE, we used the signal-normalized MSE as the validation metric.} and $R^2=0.958$. These results demonstrate that the model preserved the dominant molecular absorption features while effectively reducing noise and contamination, supporting the visual assessment presented in \autoref{fig:spectrum_cleaning}. This level of performance is not uniform across all features: weak or narrow bands associated with trace abundances may retain non-negligible errors when their signal is close to or below the effective information content of the observation (see, for instance, the O$_3$ features in the reconstructed spectra of the third row in \autoref{fig:spectrum_cleaning}). Still, it is worth noting that this limitation also affects uncontaminated transmission spectra, whose molecular detectability depends on abundance, S/N, instrumental sensitivity, and the number of observed transits \citep{linDifferentiatingModernPrebiotic2021,lustig-yaegerEarthTransitingExoplanet2023}. Moreover, we emphasize that the G-DAE should be understood as a method for mitigating contamination and recovering the dominant spectral structure, rather than as a guarantee of exact recovery for every weak feature.
\renewcommand{\thefootnote}{\arabic{footnote}}

\subsection{Quantifying uncertainties}
\label{subsec:uncertainties}

To quantify uncertainty in the reconstructed exoplanet transmission spectra from our G-DAE, we adopted a Bayesian deep learning framework. This approach aims to systematically characterize and separate the sources of uncertainty, distinguishing between two primary types: aleatoric (random) and epistemic uncertainties. Aleatoric uncertainty refers to noise inherent in the observations. It represents variability from instrumental or environmental sources \citep{kendallWhatUncertaintiesWe2017}. This form of uncertainty cannot be reduced, even with more extensive data used during model training or additional observational examples, because it reflects fundamental limitations of the measurement process, including sensor inaccuracies and stochastic perturbations in the signal. In contrast, epistemic uncertainty accounts for ignorance in the model parameters, arising from limited training data or model approximations, and can be mitigated as more data become available \citep{kendallWhatUncertaintiesWe2017,galDropoutBayesianApproximation2016}.

To estimate epistemic uncertainty without retraining the G-DAE, we employed Monte Carlo Dropout, which treats dropout layers in the neural network as a variational approximation to Bayesian inference in deep Gaussian processes \citep{galDropoutBayesianApproximation2016}. By performing multiple forward passes with dropout activated during inference (e.g., $100$ iterations), we sampled from the approximate posterior distribution over the model weights. The predictive mean was computed as the average of these reconstructions, while the variance across samples quantified the epistemic uncertainty, capturing model ignorance in regions with sparse training data or in extrapolation \citep{galDropoutBayesianApproximation2016}. This approach ensures computational efficiency while preserving fidelity to the trained model, as dropout marginalizes over the model's parameters without additional optimization, thereby providing a practical Bayesian approximation that improves predictive log-likelihood and root-mean-square error (RMSE) compared to deterministic models.

Aleatoric uncertainty was derived directly from the instrumental noise profiles of the input spectra. However, adopting the full magnitude of these observational error bars would yield overly conservative uncertainty estimates that do not reflect the G-DAE’s denoising performance. Conversely, ignoring this noise component would lead to unrealistically narrow error bars and overconfident predictions. Consequently, we defined the aleatoric uncertainty as the instrumental noise scaled by a multiplicative factor of $0.5$. This choice is heuristic but was informed by testing multiple scaling factors and selecting the value that provides a compromise between retaining the contribution of instrumental noise and avoiding overly conservative or overconfident predictions. The resulting reduced chi-square values (see next section) should therefore be interpreted primarily as diagnostics of the reconstruction residuals with respect to the assumed uncertainty model, rather than as an independent measure of goodness-of-fit or as a test for overfitting. In summary, by summing the variances of the aleatoric and epistemic components, we obtain the total predictive uncertainty \citep{kendallWhatUncertaintiesWe2017}.

\section{Model evaluation}
\label{subsec:gdae_evaluation}

To evaluate the performance of our DAE, we conducted several numerical experiments using spectra different from those used in the validation process (see \autoref{subsec:gdae_training}). We call this new set of spectra the ``evaluation set.''

To compare the reconstructed spectrum with the original uncontaminated spectrum, we used the chi-square statistic ($\chi^2$) defined by

\begin{equation}
\chi^2 = \sum_i \frac{[S(\lambda_i) - S_r(\lambda_i)]^2}{\Delta S(\lambda_i)^2},
\end{equation}
where $S(\lambda_i)$ is the clean, uncontaminated spectrum in physical units (transit depth), $S_r(\lambda_i)$ is the G-DAE-reconstructed spectrum after rescaling from normalized space, and $\Delta S(\lambda_i)$ is the wavelength-dependent uncertainty associated with the reconstructed spectrum, as returned by the G-DAE (see \autoref{subsec:uncertainties}).

We used the reduced chi-square ($\chi^2_r$) as a key metric, defined as
\begin{equation}
\chi^2_r = \frac{\chi^2}{N - p},
\end{equation}
where $N$ is the number of spectral points and $p$ is the number of fitted parameters. Since no parameters are fitted to each spectrum during inference, we set $p = 0$. This choice reflects the fact that the G-DAE does not fit physical parameters separately for each spectrum during inference. Consequently, the statistic should not be interpreted in exactly the same way as a reduced chi-square obtained from a conventional parametric fit.

A value of $\chi_r^2 \approx 1$ indicates that the reconstruction residuals are consistent with the adopted uncertainty model. In a denoising problem, however, values below unity may arise when the reconstruction residuals are smaller than the adopted predictive uncertainties, particularly for flat or weakly structured spectra. Therefore, low $\chi_r^2$ values should not be interpreted alone as evidence of perfect reconstruction, overfitting, or information leakage. Reconstruction fidelity should instead be assessed jointly through the signal-normalized MSE, $R^2$, inspection of the reconstructed spectral features and residuals, downstream retrieval performance, and controlled out-of-distribution tests.

\subsection{Denoising with realistic instrumental noise}
\label{subsubsec:gdae_evaluation_denoising}

For the first experiment, we evaluated our G-DAE using spectra degraded by \textsc{PandExo}-generated noise \citep{batalhaPandExoCommunityTool2017}. For this experiment, we simulated JWST/NIRSpec PRISM-mode time-series observations covering the spectral range 0.69--5.3~$\mu$m with the SUB512 subarray and NGROUP = 6 \citep{JDocs-NIRSpec-BOTS,JDocs-NIRSpec-Subarrays,LustigYaeger2019AJ}. We set the observation baseline to three times the transit duration (including pre- and post-transit phases), matching \exo{TRAPPIST-1}{}{e} parameters. The number of transits ranged from one to 20. For each spectrum, \textsc{PandExo} returns wavelength-dependent 1$\sigma$ uncertainties, which are important for computing the $\chi^2$ statistic (see below).

The simulation employed the PHOENIX stellar spectral energy distribution from \textsc{PandExo} using the \exo{TRAPPIST-1}{}{} parameters. For the training set, we also used the PHOENIX database, but within the context of the \textsc{TauREx} tools. Since both tools may use the spectra database differently, for instance by employing different grid interpolation procedures or parameters, the spectra used for training and testing may differ slightly, making the denoising procedure more challenging.

\begin{figure}
    \centering
    \includegraphics[width=1\linewidth]{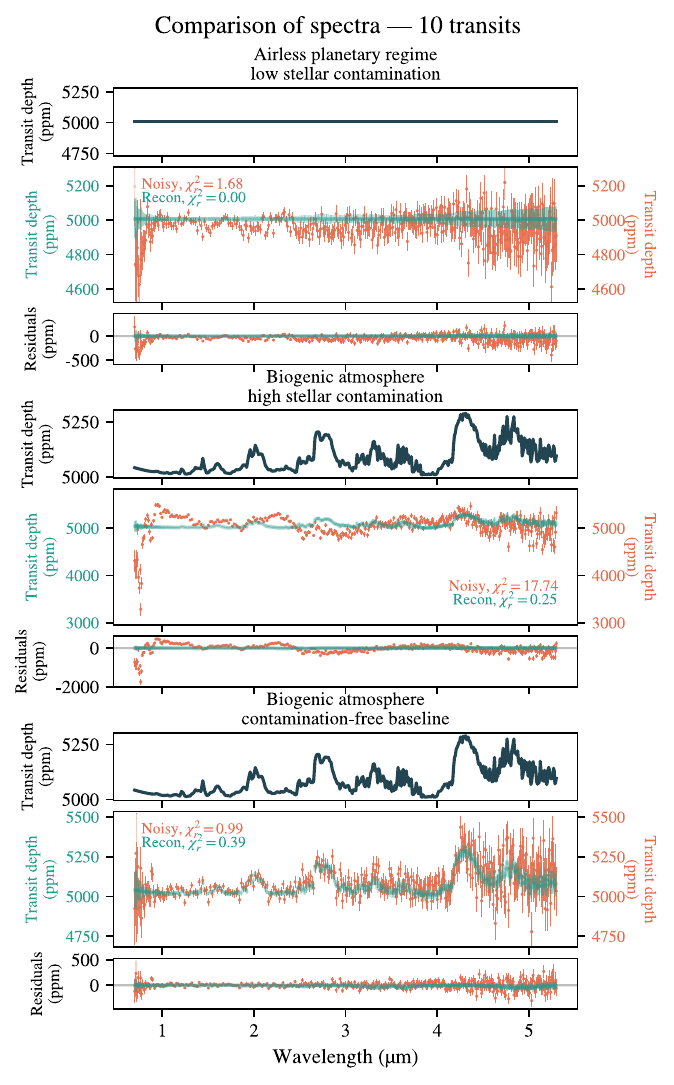}
    \caption{Comparison between noisy and contaminated input spectra (red) and G-DAE reconstructions (teal), both with error bars. The data correspond to a \exo{TRAPPIST-1}{}{e} analog with varying atmospheric compositions and stellar contamination, observed with JWST/NIRSpec over ten transits. Each case shows the original uncontaminated spectrum (top), the noisy and reconstructed spectra (middle), and the residuals of the contaminated and reconstructed signals (bottom). 
  }\label{fig:ae_pandexo_correction}
\end{figure}

\autoref{fig:ae_pandexo_correction} shows representative G-DAE reconstructions of different types of transmission spectra. In all scenarios, the application of the G-DAE reduces both contamination and noise, producing flatter residuals and much closer matches to the uncontaminated spectrum. This is reflected in lower $\chi_r^2$ values for the reconstructed signal than for the noisy case, although their absolute values depend on the adopted uncertainty scale and the complexity of the planetary spectrum.

\subsection{Quantitative assessment of model performance}
\label{subsec:gdae_quantitative_performance}

\begin{figure}
    \centering
    \includegraphics[width=1\linewidth]{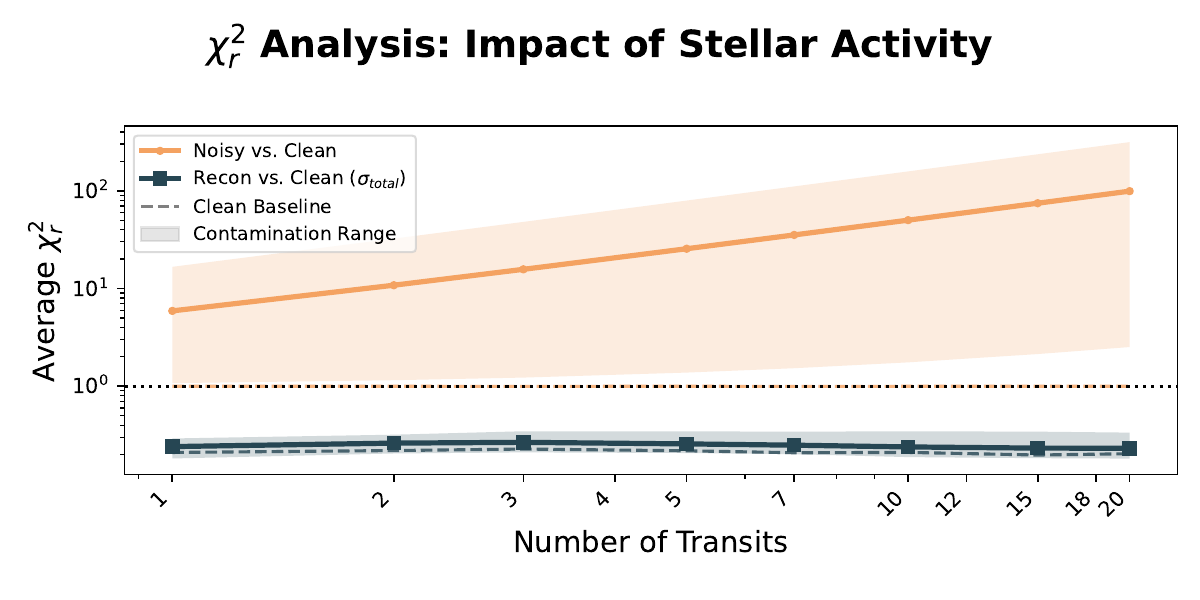}
    \caption{Average reduced chi-square $\chi^2_r$ (solid line) for G-DAE reconstructed spectra versus number of transits (noise level). The shaded band shows the range of stellar contamination used in our experiments. For comparison, the dashed line gives the $\chi^2_r$ obtained by comparing uncontaminated noisy spectra with the original spectrum.
    }
    \label{fig:chi2_analysis}
\end{figure}

To evaluate the model's performance over a wider range of conditions, we applied the G-DAE to spectra with different noise levels, i.e., assuming different numbers of transits per spectrum. The larger the number of transits, the lower the noise level. For each spectrum, we applied different levels of contamination and used the G-DAE to reconstruct the original signal. For each signal, we computed, as before, realistic instrumental noise. The results are shown in \autoref{fig:chi2_analysis}. As a side note, for producing this figure, we applied the G-DAE to a total of $3\times 10^{6}$ spectra, and it took less than 5 minutes on a workstation with an AMD Ryzen 7 5700X3D processor (eight cores and 16 threads, 3.0 GHz) to obtain the reconstructed signals. This provides an estimate of the computational efficiency of this approach.

For the aggregate cases shown, the reconstructed spectra yield reduced $\chi_r^2$ values below unity and remain relatively stable as the noise level increases; that is, as the number of transits decreases. At higher S/N, the smaller instrumental uncertainties provide more informative inputs, allowing the G-DAE to produce more precise reconstructions. At lower S/N, the denoising objective still enables the recovery of the dominant spectral structure despite the larger observational uncertainties. Taken together, these aggregate results indicate that, at each noise level, the G-DAE preserves the dominant spectral information and produces reconstructions whose residuals remain consistent with the uncertainty level of the corresponding input spectrum.

Across the explored range, the reduced $\chi_r^2$ exhibits only a weak dependence on the number of transits, suggesting that the reconstruction residuals increase approximately in proportion to the observational uncertainties. Thus, the reconstruction quality relative to the noise level remains broadly stable across the explored range.

It is worth noting that despite being trained solely with constant, wavelength-independent Gaussian noise parameterized by S/N, the G-DAE performs comparably on \textsc{PandExo}-degraded spectra—whose uncertainties vary with wavelength. This demonstrates robustness to noise-model mismatch and a useful degree of domain generalization.

\subsection{Comparative evaluation of retrieval with \textsc{POSEIDON}}
\label{subsec:GDAE_retrieval}

It is essential to evaluate whether the G-DAE provides a tangible contribution to atmospheric retrieval under conditions of stellar contamination. To this end, we designed a controlled experiment in which both retrieval strategies were applied to five independent \textsc{PandExo} noise realizations of a fixed ten-transit observational campaign. The contamination level and the stellar grid used to generate the contamination were varied, while the atmospheric truth and instrumental setup were kept fixed. This setup allows for a direct comparison of the accuracy and computational costs of two distinct strategies.

As a test case, we employed a synthetic ``CO$_2$-only'' spectrum, with N$_2$ as the background gas and CO$_2$ as the sole active absorber. The spectrum was sampled at 385 wavelength channels spanning 0.69–5.3 $\mu$m, with uncertainties generated by \textsc{PandExo} configured for JWST/NIRSpec PRISM. The reference observational campaign considered here consisted of ten transits, repeated over five independent PandExo-noise realizations. We considered one uncontaminated reference case, $(f_{\rm spot},f_{\rm fac})=(0.00,0.00)$, and two contaminated branches generated with PHOENIX- and SPHINX-based stellar grids at three increasing contamination levels: $(0.01,0.08)$, $(0.08,0.54)$, and $(0.26,0.70)$.
 Thus, the retrieval experiments encompassed both the uncontaminated case and cases affected by stellar heterogeneity.

\begin{figure}
    \centering
    \includegraphics[width=\linewidth]{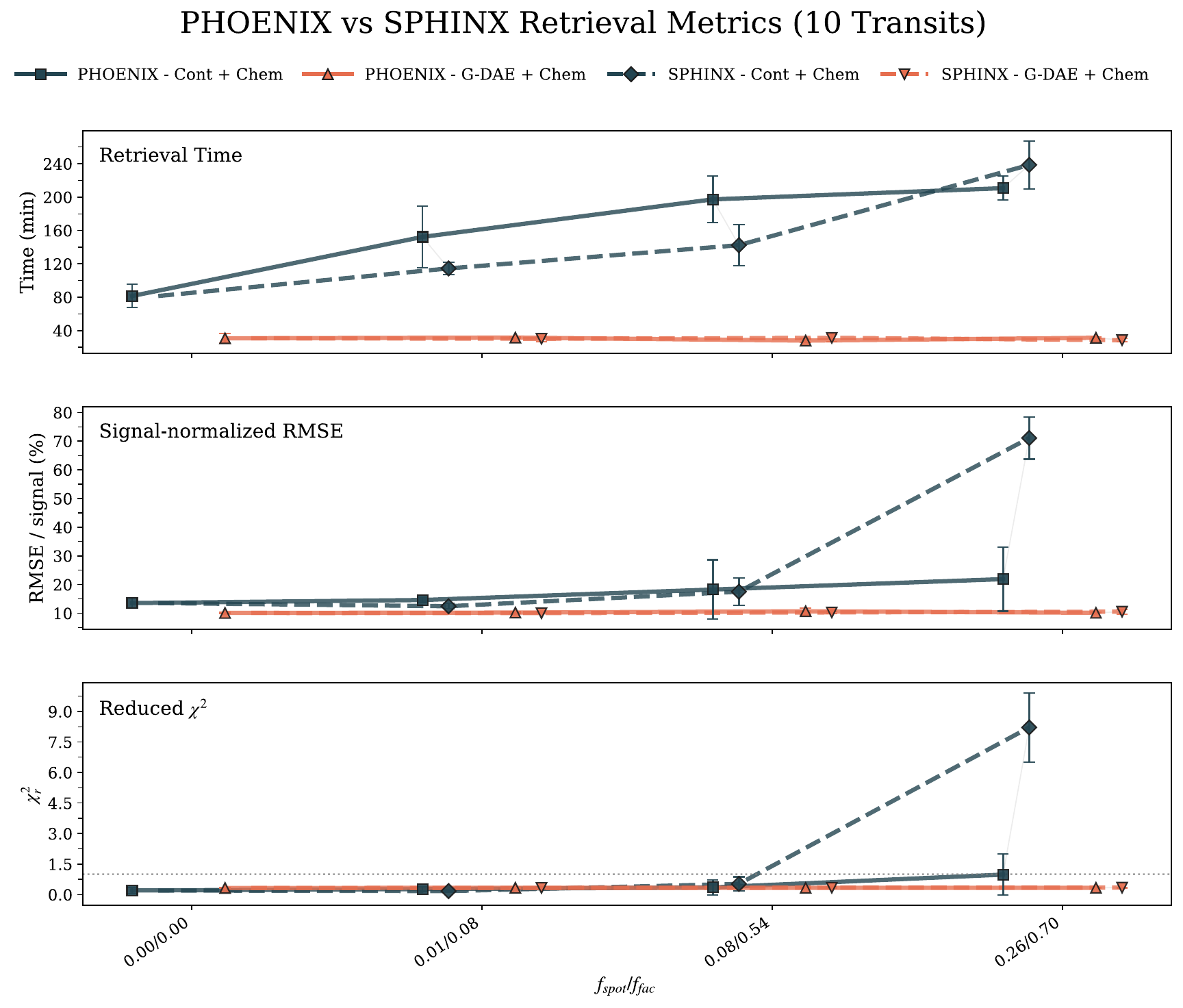}
    \caption{Comparison of retrieval strategies for a single underlying uncontaminated spectrum. Colors show methods (blue: Cont.+Chem.; red: G-DAE+Chem.), line style the stellar grid for contaminated spectra (solid: PHOENIX; dashed: SPHINX), and marker shape the branch--method combination (squares and diamonds: Cont.+Chem.; upward- and downward-pointing triangles: G-DAE+Chem.). The retrievals adopted a PHOENIX stellar grid, so the SPHINX branch probes the effect of a stellar-grid mismatch. \textbf{Top}: Wall-clock time required by the atmospheric retrieval. \textbf{Middle}: Signal-normalized RMSE, $100\,{\rm RMSE}/\Delta D_{\rm clean}$, expressed as a percentage, where $\Delta D_{\rm clean}$ is the peak-to-baseline amplitude of the uncontaminated spectrum. \textbf{Bottom}: Reduced $\chi^2$. Markers show means over five noise realizations; error bars show the $1\sigma$ dispersion.}
    \label{fig:retrieval_comparison}
\end{figure}

The results of the experiments are shown in \autoref{fig:retrieval_comparison}. The two retrieval strategies compared were: (i) Contamination plus Chemistry (Cont.+Chem.), in which stellar spots and faculae are explicitly modeled and retrieved; and (ii) G-DAE preprocessing plus atmospheric retrieval (G-DAE+Chem.), where the contaminated spectrum is first corrected with the autoencoder and then retrieved with an atmospheric forward model without explicit stellar-contamination parameters. To test the robustness of these strategies against the adopted stellar model, we carried out the benchmark with synthetic observations contaminated using either PHOENIX- or SPHINX-based stellar grids. This test is motivated by recent discussions on robust stellar-contamination corrections, which emphasize that retrievals relying on explicit stellar models can become sensitive to imperfections in the adopted stellar grid, particularly for late-type M dwarfs (RW2024).

All retrievals were conducted with \textsc{POSEIDON} using an isothermal cloud-free transmission model with 100 layers ($P_{\rm ref}=1$ bar). The state vector included the following free parameters: the $\log_{10}$ volume mixing ratios (VMRs) of H$_2$O, CH$_4$, CO$_2$, and O$_3$, the isothermal temperature, $T$, and the reference planetary radius, $R_{p,{\rm ref}}$, at 1 bar. The mixing ratio of N$_2$ was fixed by closure. The injected atmosphere contained CO$_2$ as the only active absorber. H$_2$O, CH$_4$, and O$_3$ were not included as absorbers in the input spectrum; in the retrieval, their absence was represented operationally by the lower bound of the abundance prior, $\log_{10}{\rm VMR}=-8$. The uniform priors for the atmospheric and stellar parameters are summarized in \autoref{tab:retrieval_priors}.

\begin{table}
\centering
\caption{Prior ranges used in the retrieval benchmark.}
\label{tab:retrieval_priors}
\begin{tabular}{ll}
\hline
Parameter & Range \\
\hline
$\log_{10}{\rm VMR}$(H$_2$O, CH$_4$, O$_3$) & $[-8,-1]$ \\
$\log_{10}{\rm VMR}$(CO$_2$) & $[-5,-1]$ \\
$T$ [K] & $[200,400]$ \\
$R_{p,{\rm ref}}$ & $[0.85,1.15]\times R_{p,\mathrm{true}}$ \\
$f_{\rm spot}$ & $[0,0.26]$ \\
$f_{\rm fac}$ & $[0,0.70]$ \\
$T_{\rm phot}$ [K]& $[0.9,1.1]T_s$ \\
$T_{\rm spot}$ [K]& $[0.8,0.9]T_s$ \\
$T_{\rm fac}$ [K]& $[T_s, T_s+150\,{\rm K}]$ \\
\hline
\end{tabular}
\tablefoot{Here, $T_s$ denotes the adopted stellar effective temperature.}
\end{table}

For the explicit stellar-contamination retrievals (Cont.+Chem.), the wavelength-dependent contamination factors, $\epsilon_\lambda$ (see \autoref{eq:epsilon}), were computed within the standard PHOENIX-based stellar framework implemented in \textsc{POSEIDON}. In this setup, \textsc{POSEIDON} interpolates within the adopted stellar grid rather than restricting the fit to discrete grid nodes. The retrievals were sampled with \textsc{MultiNest} \citep{ferozMultiNestEfficientRobust2009, buchnerXraySpectralModelling2014}, with $N_{\rm live}=500$, parallelized with the Message Passing Interface (MPI) across 12 processors. We assumed Gaussian likelihoods, with wavelength-dependent $1\sigma$ uncertainties calculated with \textsc{PandExo}. All retrieval computations were carried out on the same workstation used for model training. 

As a stochastic nested-sampling algorithm, \textsc{MultiNest} explores the parameter space subject to progressively increasing likelihood constraints. Although the formal parameter space and sampling configuration remain unchanged across cases, increasing stellar heterogeneity can alter the geometry and degeneracies of the posterior, reducing the efficiency of drawing new live points and thereby increasing the wall-clock time. Consequently, the runtime depends not only on the number of data points but also on the sampling difficulty induced by the posterior geometry.

In the top panel of \autoref{fig:retrieval_comparison}, explicit modeling of stellar contamination (Cont.+Chem.; blue curves) was the most computationally expensive strategy. By contrast, G-DAE+Chem. consistently had much lower runtimes—roughly one order of magnitude faster across the benchmark, and slightly exceeding this margin in the most contaminated cases. For the signal-normalized RMSE (middle panel in \autoref{fig:retrieval_comparison}) and the reduced $\chi^2$ (bottom panel), the relevant comparison is between matched and mismatched stellar-contamination models. When the contaminated spectra are generated with the same stellar family adopted by the retrieval framework (PHOENIX), Cont.+Chem. and G-DAE+Chem. yield broadly comparable performance. However, when the spectra are generated with SPHINX while the retrieval still relies on the standard PHOENIX-based stellar prescription, Cont.+Chem. degrades more strongly as the contamination level increases. By contrast, G-DAE+Chem. remains comparatively stable across both stellar-grid branches. Taken together, these findings show that G-DAE preprocessing provides the best compromise between computational efficiency and robustness, especially when the stellar contamination departs from the spectral grid assumed by the explicit retrieval model.

\subsection{Recovered atmospheric parameters}

\begin{figure}
    \centering
    \includegraphics[width=1\linewidth]{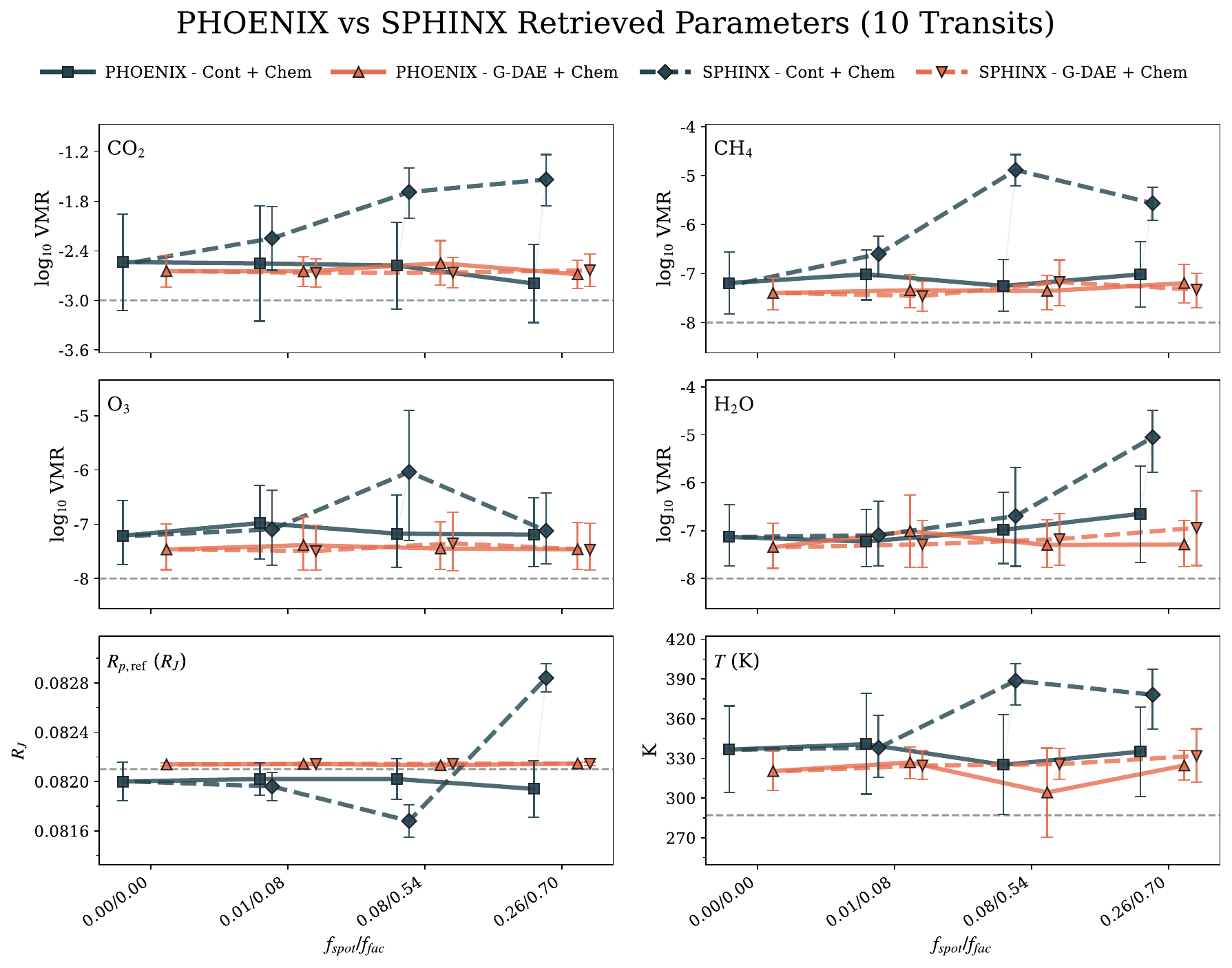}
    \caption{Retrieved atmospheric parameters for the two retrieval strategies. The dashed gray horizontal line marks the ground-truth value used to generate the synthetic spectrum.}
    \label{fig:vmr_vs_contam}
\end{figure}

To inspect how the retrieval strategy affects the inferred atmospheric composition, we analyzed the retrieved VMR and nuisance parameters obtained with the various strategies described in the previous section. The results are summarized in \autoref{fig:vmr_vs_contam}. The panels detail the inferred abundances ($\log_{10}$ VMR) for the injected active absorber (CO$_2$) as well as the inactive molecules (CH$_4$, O$_3$, H$_2$O), alongside the reference planetary radius, $R_{p,\mathrm{ref}}$, and the isothermal temperature, $T$. To capture the full scope of uncertainty in these measurements, the error bars combine the intrinsic width of the retrieval posterior with the realization-to-realization scatter across the five simulated observations.

In this experiment, the lower bound of the prior ($10^{-8}$) is taken to represent the absence of the molecule (see \autoref{tab:retrieval_priors}). Because a retrieval yields a posterior distribution broadened by instrumental noise, the median may lie slightly above the $-8$ level in $\log_{10}$ VMR even when there is no signal; this shift is a statistical effect of posterior broadening and should be interpreted as non-detection (an upper limit consistent with the prior).

Overall, G-DAE+Chem. recovers planetary parameters most consistently across both stellar grids. While both pipelines perform similarly well when the injected contamination matches the retrieval model (PHOENIX) (RW2024), their behavior diverges under a grid mismatch (SPHINX). In this mismatched scenario, Cont.+Chem. systematically biases the recovered values for CO$_2$, CH$_4$, H$_2$O, $R_{p,\mathrm{ref}}$, and $T$, with the drift worsening as stellar contamination increases. By contrast, G-DAE+Chem. remains substantially more robust, generally keeping inactive molecules near the non-detection regime with appropriately broadened uncertainties, although localized deviations persist in the most contaminated regime (specifically in $T$ and one H$_2$O solution). Ultimately, G-DAE's data-driven preprocessing provides superior reliability by avoiding the need to force the retrieval algorithm through a rigidly mismatched stellar grid.

\section{Other planetary cases}
\label{sec:other_planets}

The case of terrestrial planets with potential biosignatures around M dwarfs was first explored in  \citet{duque-castanoMachineassistedClassificationPotential2025} and is further developed in the present work.  However, these techniques may be particularly useful for larger planets, where transmission signals are generally stronger and observationally more accessible. In particular, sub-Neptunes and giant planets provide a promising regime in which to test the performance of the methods developed in this work.

In the following sections, we extend our workflow to the case of sub-Neptunes, a class of exoplanets with radii between 1 and 4 Earth radii that, according to population studies (see \citealt{parcSuperEarthsSubNeptunesObservational2024} and references therein), represent a prevalent type of planet in the Galaxy. These worlds offer a unique opportunity for in-depth characterization of intermediate-sized exoplanets' interiors and atmospheres \citep{huWaterrichInteriorTemperate2025, liuHydrocarbonHazesTemperate2026,guzman-mesaChemicalDiversityAtmospheres2022, parcSuperEarthsSubNeptunesObservational2024} and could also offer interesting possibilities for the search of biosignatures (see \citealt{madhusudhanHyceanParadigmSearch2024} and references therein). 

\subsection{A model case for sub-Neptunes}
\label{subsec:model_subneptunes}

As we did for the case of terrestrial planets with \exo{TRAPPIST-1}{}{e}, we selected a model case to analyze sub-Neptunes' transmission spectra. Given the challenges of interpreting its spectrum and the recent interest in the planet, we used the potential Hycean planet \exo{K2}{-18}{b} \citep{madhusudhanCarbonbearingMoleculesPossible2023} as a model case. 

\exo{K2}{-18}{b} is a sub-Neptune orbiting within the habitable zone of its host star, an earlier-type red dwarf \citep{huWaterrichInteriorTemperate2025} (see \autoref{tab:k218b} for a summary of the planetary and stellar properties).  With a mass of 7.2 Earth masses and a radius of $\sim$2.5 Earth radii, the location of \exo{K2}{-18}{b} within its star's habitable zone suggests the potential for liquid water and, consequently, some prospects of habitability \citep{tsiarasWaterVapourAtmosphere2019, madhusudhanHyceanParadigmSearch2024, madhusudhanNewConstraintsDMS2025}. This has made \exo{K2}{-18}{b} a prime candidate for scrutinizing the atmospheric composition and internal structure of habitable sub-Neptunes, especially around M dwarfs (see e.g., \citealt{liuHydrocarbonHazesTemperate2026}).

\begin{table}
\centering
\caption{System parameters of \exo{K2}{-18}{b} used in the sub-Neptune experiments.}\label{tab:k218b}
\begin{tabular}{l l}
\hline
\multicolumn{2}{c}{\textbf{Star (\exo{K2}{-18}{})}} \\
\hline
Radius ($R_\odot$) & 0.468 \\
Mass ($M_\odot$) & 0.495 \\
$T_{\text{eff}}$ (K) & 3500 \\
\\
\hline
\multicolumn{2}{c}{\textbf{Planet (\exo{K2}{-18}{b})}} \\
\hline
Radius ($R_\oplus$) & 2.461 \\
Mass ($M_\oplus$) & 7.2 \\
Semi-Major Axis (au) & 0.1429 \\
\hline
\end{tabular}
\tablebib{\citet{howardPlanetMassesRadii2025}.}
\end{table}

The JWST has significantly advanced the characterization of \exo{K2}{-18}{b}'s atmosphere, revealing the presence of several chemical signatures \citep{madhusudhanHyceanParadigmSearch2024, madhusudhanCarbonbearingMoleculesPossible2023}. Specifically, observations have revealed prominent spectroscopic signatures of methane and have motivated continued investigation of other atmospheric species, including carbon dioxide \citep{madhusudhanHyceanParadigmSearch2024,liuHydrocarbonHazesTemperate2026}. 

Although, as has been extensively argued in this work and the references herein, stellar contamination from the host star can often interfere with transit spectroscopy, \exo{K2}{-18}{b}'s M dwarf host star offers a unique advantage due to its relatively quiet nature compared to more active late-type M dwarfs \citep{madhusudhanHyceanParadigmSearch2024}. However, accurate atmospheric retrieval for planets orbiting such stars still requires accounting for potential stellar activity, which can introduce variability into spectroscopic data \citep{rackhamEffectStellarContamination2023}. 

In summary, although \exo{K2}{-18}{b} is an interesting scientific target in its own right, the purpose of using this planet as a model case does not imply that our conclusions are restricted to this particular planet. Our main aim is to use the well-known properties of \exo{K2-18}{}{} and one of its planets to synthesize planetary analogs for training our algorithms and to gain insights into their performance when decontaminating sub-Neptune transmission spectra. 

\subsection{Atmospheric composition}
\label{subsec:subneptunes_atmosphere}

To model the atmosphere of our mock sub-Neptunes, we adopted the atmospheric parameterization of the Atmospheric Big Challenge (ABC) Database presented as part of the ESA-Ariel Data Challenge NeurIPS 2022 \citep{changeatESAArielDataChallenge2023}. Synthetic spectra were generated using the \href{https://ascl.net/2504.022}{\textsc{MultiREx}} framework \citep{duque-castanoMachineassistedClassificationPotential2025} with stellar and planetary parameters taken from \citet{howardPlanetMassesRadii2025}. 

All mock planets were modeled with well-mixed, cloud-free atmospheres and free molecular abundances, at least above a base pressure of 10 bar. As in the case of terrestrial planets, we assumed an isothermal atmospheric profile here. In contrast to the terrestrial planet case, we used a finer training grid of atmospheric temperatures, ranging from 250 K to 450 K in 50 K increments. For the top pressure boundary, we used a fiducial value of $10^{-3}$ bar. The fill gases are those of primary atmospheres of H$_2$ and He, and hence, for modeling the transmission spectrum, we included collision-induced absorption (CIA) features for H$_2$-H$_2$ and H$_2$-He \citep{gordonHITRAN2020MolecularSpectroscopic2022}.

Following the ARIEL Data Challenge specifications, we considered four molecular species with the following $\log_{10}$ VMR ranges: CH$_4$ from $-8$ to $-1$, CO$_2$ from $-10$ to $-1$, H$_2$O from $-7$ to $-2$, and NH$_3$ from $-7$ to $-2$. The molecular absorption cross-sections were taken from ExoMol \citep{chubbExoMolOPDatabaseCross2021,tennysonExoMolDatabaseMolecular2016}. We excluded CO because, at the noise levels of our numerical experiments, its bands mostly overlap with those of CO$_2$, and their strengths are too low to justify the additional computational cost of generating additional synthetic spectra.

As we did in the terrestrial planet model case, we also included in the sample of mock sub-Neptunes, on the one hand, planets having some or most of the gases completely absent, and, on the other hand, planets with flat transmission spectra. Conceptually, the latter planets would correspond to sub-Neptunes with very high, opaque clouds. Numerically, those flat spectra are simply signals of equal strength (the nominal radius of the planet) at all sampled wavelengths. 

\subsection{Stellar contamination}
\label{subsec:subneptunes_stellar_contamination}

Since \exo{K2}{-18}{}\ (the host star of our sub-Neptune model) is less active than \exo{TRAPPIST-1}{}{} (our terrestrial planet baseline), we adapted the stellar contamination parameters used to train our algorithms and used four equidistant values for the spot covering fraction, $f_\mathrm{spot}$ (from 0.0 to 0.3), and four for the facula covering fraction, $f_\mathrm{fac}$ (from 0.0 to 0.4). For the training grid, we sampled three photospheric temperatures, $T_\mathrm{phot} = 3400$, $3500$, and $3600$ K, spanning the range estimated for \exo{K2}{-18}{} (see \autoref{tab:k218b}). The temperatures of spots and faculae followed the ansatz from \citet{rackhamTransitLightSource2018}, with  $T_\mathrm{spot} = 0.86 \times T_\mathrm{phot}$ and $T_\mathrm{fac} = T_\mathrm{phot} + 100$ K. 

\subsection{The sub-Neptunes dataset}
\label{subsec:subneptunes_dataset}

Using the previously chosen model parameters, the dataset for sub-Neptunes comprised $1{,}058{,}416$ spectra = ($66{,}150$ atmospheric compositions + one flat model) $\times$ 16 contamination scenarios.  Additionally, we augmented the database with three transit configurations (noise levels), namely one, two, and four total transits. For each transit configuration, we computed the corresponding instrumental noise individually.  Technically, a fourth transit configuration, corresponding to a clean spectrum, which can be regarded as an infinite number of transits, was also provided to the algorithm during the training phase. The total volume of the dataset was $7{,}793{,}440$ pairs of spectra (uncontaminated and contaminated), divided into 80\% ($\sim$6.2M) for model training and 20\% ($\sim$1.5M) for testing.

\subsection{Model validation and evaluation}

\begin{figure}[ht!]
    \centering    
    \includegraphics[width=1\linewidth]{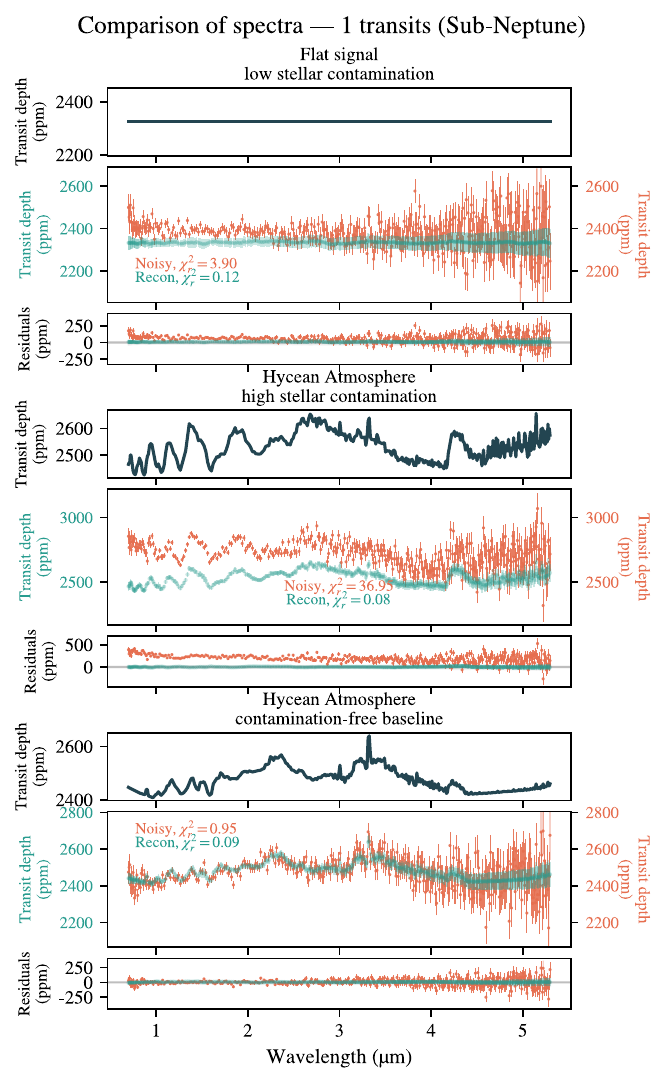}
    \caption{Comparison between noisy and contaminated input spectra (red markers) and the reconstructions performed by our G-DAE (teal markers), both with error bars, for a \exo{K2}{-18}{b} analog (Sub-Neptune).
    We consider three scenarios: a flat signal (top row), a Hycean atmosphere with high stellar contamination (middle row), and the same Hycean atmosphere without stellar contamination (bottom row), observed with JWST NIRSpec using a single transit.
    The panel structure is analogous to \autoref{fig:ae_pandexo_correction}.}
    \label{fig:subneptune_reconstruction}
\end{figure}

As we did in \autoref{subsubsec:gdae_evaluation_denoising} and \autoref{fig:ae_pandexo_correction}, to validate the performance of the sub-Neptune G-DAE, we first performed a simple visual inspection of reconstructed spectra at different composition and contamination levels. Three examples of the results returned by the G-DAE trained on the sub-Neptunes dataset are shown in \autoref{fig:subneptune_reconstruction}. A simple visual inspection, confirmed by the value of the $\chi^2_r$ statistic, shows that the G-DAE is very successful at decontaminating sub-Neptune transmission spectra. This confirms that the power of the G-DAE is not constrained to the particular chemical composition, size, or stellar type we tested in our first numerical experiments.

A more in-depth, quantitative assessment of the algorithm's performance (model validation) on the test dataset yielded a signal-normalized MSE of $0.0109\%$, more than an order of magnitude lower than the corresponding metric in the terrestrial-planet case (see \autoref{subsec:gdae_training}). This is explained by the significantly higher S/N of the spectra for larger planets. On the other hand, the test coefficient of determination in this case is $R^2=0.992$, which again is significantly better, on the scale of this particular metric, than in our first numerical experiment.

\begin{figure}[t!]
    \centering   
    \includegraphics[width=1\linewidth]{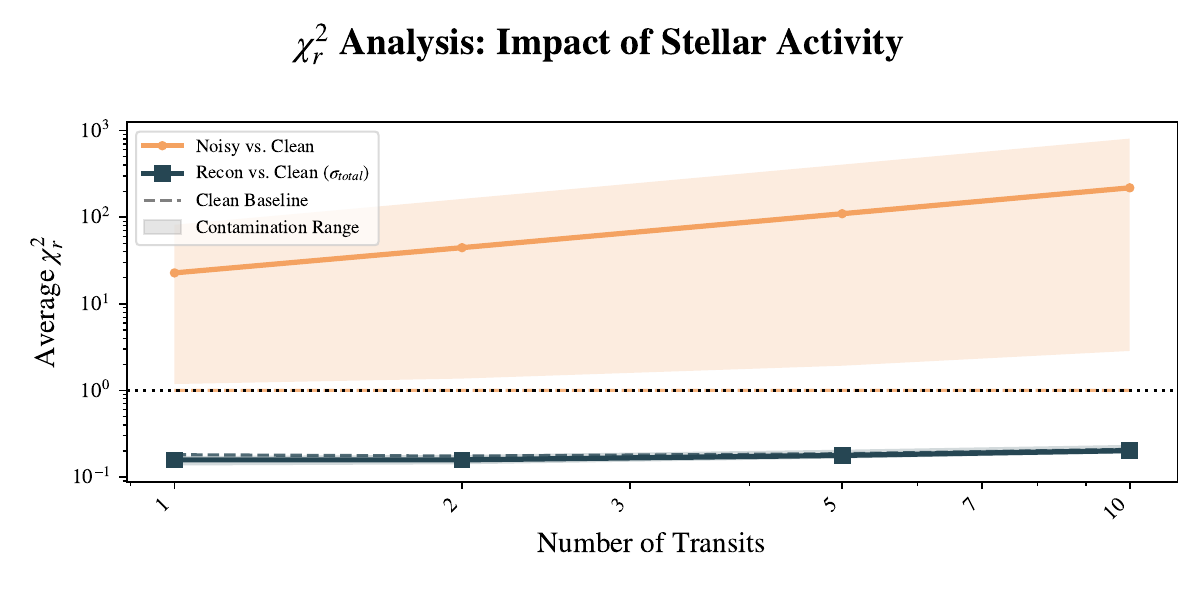}
    \caption{Average reduced chi-square $\chi^2_r$ (continuous line) for G-DAE reconstructed spectra of sub-Neptune analogs as a function of the number of transits (noise level). This figure is similar to \autoref{fig:chi2_analysis}.}
    \label{fig:subneptune_chi2r}
\end{figure}

Lastly, we analyzed how G-DAE performance depends on the number of transits, which determines the instrumental noise level and hence the S/N. The results for the sub-Neptune case are presented in \autoref{fig:subneptune_chi2r}; the corresponding terrestrial-planet results are shown in \autoref{fig:chi2_analysis} and discussed in \autoref{subsec:gdae_quantitative_performance}.

Although the conclusions in this case are very similar to those for the terrestrial-planet case discussed above, the dependence of $\chi^2_r$ for the sub-Neptune G-DAE shows striking similarities to that of the terrestrial-planet G-DAE (see \autoref{fig:chi2_analysis}). This similarity confirms that the behavior of the autoencoders is mostly independent of atmospheric model parameters and even signal strength.

\section{Discussion and future directions}
\label{sec:discussion}

After designing and successfully testing our G-DAEs for the particular conditions of our experimental setup, several questions remain open. In this section, we assess the key concerns and caveats of our approach.

The first concern is whether, as is common in machine-learning applications, the neural networks we trained are overly specific (overfitted) to the types of planets used in the training process. More broadly, this raises the issue
of whether this type of algorithm is sufficiently general to analyze real transmission spectra.  A key characteristic of our method that reduces the risk of overfitting is the high diversity of the training datasets. As explained in \autoref{subsec:gdae_training} and \autoref{subsec:subneptunes_dataset}, although the number of molecular species is relatively small, their mixing ratios span several orders of magnitude, yielding a large and heterogeneous collection of spectra. In addition, the reference-depth representation reduces the dynamic range associated with absolute transit-depth levels and allows the model to focus primarily on spectral morphology.

The effect of stellar effective temperature, $T_\mathrm{eff}$, is more complex. Assuming the same spot and facula covering fractions, a cooler or hotter star produces stellar contamination with different strengths and at different wavelengths. A G-DAE trained for a given $T_\mathrm{eff}$ could be ``confused'' when fed with a transmission spectrum of a star with a different temperature. We tested the effect of stellar temperature and found that, over a range of a few hundred kelvins, stellar contamination does not vary significantly with wavelength. As a result, the effect of, for instance, a slightly lower temperature is interpreted by the algorithm as a slightly lower contamination. If, on the other hand, the effective temperature of the true star is much higher than that used in the training dataset, the true contamination, due to astrophysical effects, could also be much lower; as a result, the G-DAE reconstructs the data as if no contamination were present. 

We took a first step in this direction by repeating the terrestrial benchmark with contamination generated from both PHOENIX and SPHINX stellar grids. The qualitative ranking of the retrieval strategies remains unchanged, but the explicit stellar-contamination retrieval becomes significantly more sensitive when the contamination is generated with a stellar grid different from the one assumed by the retrieval model. This result reinforces the idea that grid mismatch is not a minor technical detail but a realistic source of bias in stellar-contamination-aware retrievals.

Another relevant question is whether the observed spectrum contains molecular species that were not part of the training set. To explore this scenario, we carried out an $N+1$ experiment: we trained a G-DAE on sub-Neptune synthetic spectra that deliberately exclude a new molecule, for example, DMS, and then used this G-DAE to reconstruct the spectrum of a test planet whose atmosphere does include DMS. This experiment is detailed in \autoref{sec:nplus1_dms}. Because the G-DAE is limited by the chemical library used in training, its reconstructions cannot perfectly match the true $N+1$ spectra. Yet its failure mode is highly informative. When the DMS signal is weak—due to trace abundances or noise-dominated data (e.g., a single transit)—the network effectively masks the feature, preserving a good fit ($\chi_r^2 < 1$) with high confidence. As the DMS abundance and S/N increase, the unexpected signature strengthens, and instead of forcing a match to a known gas, the G-DAE proportionally inflates its epistemic uncertainty in the mismatched wavelength band.

This experiment also highlights the complementary roles of $\chi_r^2$ and epistemic uncertainty. A weak unmodeled signal may be attenuated while remaining consistent with the adopted predictive uncertainties, whereas, in our experiment, stronger discrepancies were accompanied by an increased epistemic contribution within the affected spectral region. Thus, the G-DAE provides both a decontaminated spectrum and a warning indicator for sufficiently strong departures from its learned spectral domain. Moreover, by passing these enlarged error bars to the retrieval algorithm, the G-DAE provides an automated diagnostic warning that relaxes the likelihood constraints, preventing overly confident, incorrect chemical fits. We verified that as long as the observed chemistry remains within the learned chemical space, epistemic uncertainties remain much lower than instrumental noise. Consequently, a sudden spike in epistemic variance serves as a highly targeted indicator of unknown species or missing physics.

The autoencoder is trained specifically to remove contamination, not to detect the signatures of particular molecular species. The G-DAEs are not intended to replace retrieval algorithms, which are intrinsically designed to handle chemical diversity. This architectural choice fundamentally distinguishes our approach from other machine learning (ML) applications in exoplanetary science. While recent literature employs techniques such as random forests, convolutional neural networks, or normalizing flows to accelerate or replace the Bayesian atmospheric retrieval step \citep{nixonAssessmentSupervisedMachine2020,ardevolmartinezConvolutionalNeuralNetworks2022,ardevolmartinezFlopPITyEnablingSelfconsistent2024}, our G-DAE operates strictly as a standalone preprocessing algorithm. By decoupling decontamination from retrieval, we avoid the heavy model dependence of end-to-end ML retrievals. Furthermore, Gaussian-process retrieval frameworks such as that of \citet{espinozaJWSTTSTDREAMSNIRSpec2025a} have been developed to model and marginalize visit-dependent stellar signals, a physically motivated strategy when stellar heterogeneities evolve between transits. However, these frameworks rely on a temporal decomposition in which signals that vary between visits are attributed to the star, whereas signals that persist across visits are attributed to the planet. Consequently, persistent stellar heterogeneities may remain degenerate with atmospheric features. By contrast, our G-DAE addresses the problem at the level of individual spectra and does not explicitly exploit temporal information across visits. Combining multiple visits before reconstruction may improve the S/N, but it implicitly assumes that the stellar contamination remains sufficiently stable across those observations. Processing visits separately preserves their individual contamination states, but the present network does not explicitly model their temporal evolution. Nevertheless, the reconstruction tests show that the G-DAE remains effective over the range of S/N explored, including low-S/N cases corresponding to only a few transits (\autoref{fig:chi2_analysis} and \autoref{fig:subneptune_chi2r}).

A few caveats should be noted regarding real-world application. By design, the input and output signals of our G-DAE are normalized using a reference transit depth, $D_{\rm ref}$, associated with the reference planetary radius, $R_{p,\rm ref}$, adopted in the synthetic forward model. The present G-DAE was not designed to marginalize over uncertainties in $R_{p,\rm ref}$ or to handle arbitrary shifts in the absolute transit-depth level caused by variations in this reference radius. In many observational cases, a useful estimate of the reference planetary radius may be obtained from broadband transit photometry, particularly when the observations are conducted in bands expected to be less affected by stellar contamination.

Several avenues for future work remain. First, while our current model assumes static heterogeneity, stellar contamination in real systems may vary from one transit to the next. Second, our analysis has focused exclusively on the JWST NIRSpec PRISM configuration; exploring other instrumental setups or entirely different observatories is a necessary next step. Finally, expanding the training grids to incorporate non-equilibrium chemistry will further extend the G-DAE's applicability to a wider range of planetary environments.

\section{Summary and conclusions}
\label{sec:summary_and_conclusions}

Stellar contamination is one of the most limiting factors in transmission spectroscopy for exoplanets of all sizes. It is particularly important for exoplanets orbiting active M dwarfs (which are abundant and have yielded particularly interesting targets for astrobiological investigations), and its deleterious effects could be independent of signal strength (planetary size). 

In this work, we present a novel neural-network-based methodology for decontaminating transmission spectra from noisy and/or contaminated signals. For that purpose, we designed and optimized the architecture of the G-DAE, a multilayer, mirror-structured neural network designed to compress and capture the main features of a planetary transmission spectrum. 

We trained two analogous but differently optimized G-DAEs. One of them specialized in the transmission spectra of terrestrial planets with atmospheres of astrobiologically relevant compositions, and the other in decontaminating the signal of sub-Neptunes. For the first one, we trained the network using planetary analogs of \exo{TRAPPIST-1}{}{e}, and for the second one, we used the sub-Neptune \exo{K2}{-18}{b} as the model planet. Although trained with very specific planetary parameters, our results suggest that the performance of the G-DAEs is largely independent of stellar and planetary parameters.

A visual inspection of the reconstructed transmission spectra (see \autoref{fig:ae_pandexo_correction} and \autoref{fig:subneptune_reconstruction}) shows that, almost regardless of noise and stellar contamination levels, the networks reconstruct the spectra with high fidelity. They learned the general features of theoretical spectra and applied their training to denoise realistic signals. A more extensive evaluation, using different metrics and millions of synthesized noisy and contaminated spectra, supports these conclusions.

While the G-DAE is a preprocessor and not a retrieval replacement, its ability to decontaminate spectra significantly improves downstream atmospheric characterization. As demonstrated by our terrestrial benchmark, utilizing G-DAE-corrected spectra reduces the computational cost of complex atmospheric retrievals by an order of magnitude or more (see \autoref{fig:retrieval_comparison}). More importantly, under strong stellar contamination—and especially when the true stellar physics departs from the models assumed by traditional retrievals—working with G-DAE-reconstructed spectra provides greater robustness and prevents the algorithm from confidently converging on incorrect atmospheric parameters (see \autoref{fig:vmr_vs_contam}).

\section*{Data availability}

All data required to replicate the results and generate the figures presented in this work, including the {\tt Jupyter} notebooks used to train and test the models, are available in the public {\tt GitHub} repository  \url{https://github.com/D4san/gdaespec}. The trained G-DAE parameters, provided in {\tt .keras} format, are available in the {\tt model-parameters/} folder within the repository. We have provided a Jupyter notebook in the same folder to illustrate the workflow for applying the trained G-DAE to a test case.

\begin{acknowledgements}

A significant fraction of the results in this work were almost impossible to obtain in the span of a human life without the use of open-source, general-purpose {\tt Python} packages, including 
\  {\tt Matplotlib} \citep{Matplotlib2007}, {\tt Numpy} \citep{Numpy2020}, {\tt Scipy} \citep{virtanenSciPy10Fundamental2020}, {\tt Scikit-learn} \citep{pedregosaScikitlearnMachineLearning2011}, {\tt Tensorflow} \citep{tensorflow2015-whitepaper}, {\tt mpi4py} \citep{dalcin2021mpi4py} and {\tt pandas} \citep{mckinney_proc_scipy_2010}. We are also grateful to the teams of scientists and developers who created {\tt TauREx} \citep{al-refaieTauREx3Fast2021} and {\tt POSEIDON} \citep{macdonaldPOSEIDONMultidimensionalAtmospheric2023}, which have enabled transmission spectroscopy for everyone. 
\\
We sincerely acknowledge the referee for their insightful comments and valuable suggestions, especially for proposing additional aspects for evaluating and training our models, which significantly enhanced their robustness.
\\
This work was supported by the Universidad de Antioquia through research project 2026-89170, ``Development of deep-learning models for reducing noise and stellar contamination in exoplanet transmission spectra.''

\end{acknowledgements}

\bibliographystyle{bibtex/aa}

\bibliography{bibtex/references}

\begin{appendix}
\nolinenumbers
\section{Machine learning for transmission spectroscopy}
\label{app:ML}

Machine-learning methods have increasingly permeated the analysis of exoplanet spectra. However, the bulk of this literature has focused on accelerating, approximating, or entirely replacing the atmospheric retrieval process, rather than cleaning the observed spectrum prior to analysis. 

Previous studies have employed random forests, convolutional neural networks, Bayesian neural networks, and normalizing flows to infer atmospheric properties directly from the light we receive, or to amortize posterior estimation \citep{nixonAssessmentSupervisedMachine2020,ardevolmartinezConvolutionalNeuralNetworks2022,vasistNeuralPosteriorEstimation2023,ardevolmartinezFlopPITyEnablingSelfconsistent2024}. In parallel, unsupervised methods have emerged for exploratory analysis and anomaly detection in vast spectroscopic datasets \citep{matchevUnsupervisedMachineLearning2022,forestanoSearchingNovelChemistry2023}. More recently, deep generative models—such as conditional variational autoencoders—have been deployed as surrogate forward models, enabling rapid spectral synthesis and data augmentation conditioned on specific planetary and stellar parameters \citep{mukherjeeConditionalVariationalAutoencoder2025}. A comprehensive compilation of ML architectures applied to atmospheric characterization is provided in Table 1 of \citet{duque-castanoMachineassistedClassificationPotential2025}.

In this work, we have focused on a different objective: designing and training specialized neural networks (autoencoders) to mitigate stellar contamination and recover the intrinsic planetary signal before standard retrieval algorithms take over. The choice of this specific architecture is motivated by its success in other astronomical domains. In recent years, autoencoders have found multiple applications within the astronomical community, demonstrating their effectiveness across diverse observational contexts and data types \citaeg{see e.g.,}{bartlettNoiseReductionSingleshot2023a, scourfieldDenoisingGalaxyOptical2023a}. Within the field of stellar spectroscopy, \cite{sedaghatStellarKaraokeDeep2023a} and \cite{kjaersgaardTAUNeuralNetwork2023} proposed an autoencoder-based approach designed to autonomously identify and eliminate telluric contamination present in stellar spectra. This methodology effectively distinguishes atmospheric absorption characteristics from stellar signatures, achieving this separation without depending on synthetic atmospheric models or specific details of the observational conditions. In exoplanetary transmission spectroscopy, the connection between stellar contamination and machine learning has begun to emerge in the recent literature. On the one hand, ML-based approaches have been proposed to identify or discriminate potential stellar contamination in exoplanet transmission observations, particularly in the context of target assessment and multi-color photometric diagnostics \citep{ying-pingDiscriminationStellarContamination2025b}. On the other hand, the 1st Ariel Machine Learning Challenge\footnote{https://arielmission.space/index.php/data-challenges/} showed that ML techniques, including deep-learning models, can correct spot-induced distortions in simulated transit light curves, highlighting the promise of these methods for activity-related systematics in exoplanet data. At the same time, the challenge also showed that competitive performance can be achieved by more interpretable, feature-based approaches, underscoring that strong physical structure remains important for robust generalization \citep{nikolaouLessonsLearned1st2023}. 

In other areas, Gaussian-process retrieval frameworks such as that of \citet{espinozaJWSTTSTDREAMSNIRSpec2025a} have been used to model and marginalize visit-dependent stellar signals, making them particularly suitable for contamination that varies between transits. Their temporal decomposition nevertheless leaves persistent stellar heterogeneities potentially degenerate with the planetary atmospheric signal.

These challenges underscore the need for a methodology that captures the physical structure of the contamination without relying strictly on multi-epoch temporal variability. In the domain of exoplanetary atmospheric characterization, our team has previously employed DAEs as a core component of an ensemble learning framework designed to infer the presence of bioindicators in low-S/N spectra \citep{duque-castanoMachineassistedClassificationPotential2025}. While those DAEs exhibited a commendable capability to reconstruct salient spectral features, that investigation was not primarily oriented toward comprehensive denoising or high-fidelity removal of stellar contamination. For rigorous and detailed analyses of exoplanetary spectra, a more versatile, informative, and scalable DAE design is required—one capable of cleaning individual spectra based on the intrinsic spectral signatures of the TLS. 

\section{Autoencoders for denoising transmission spectra}
\label{sec:autoencoders}

\begin{figure}
    \centering
    \includegraphics[width=1\linewidth]{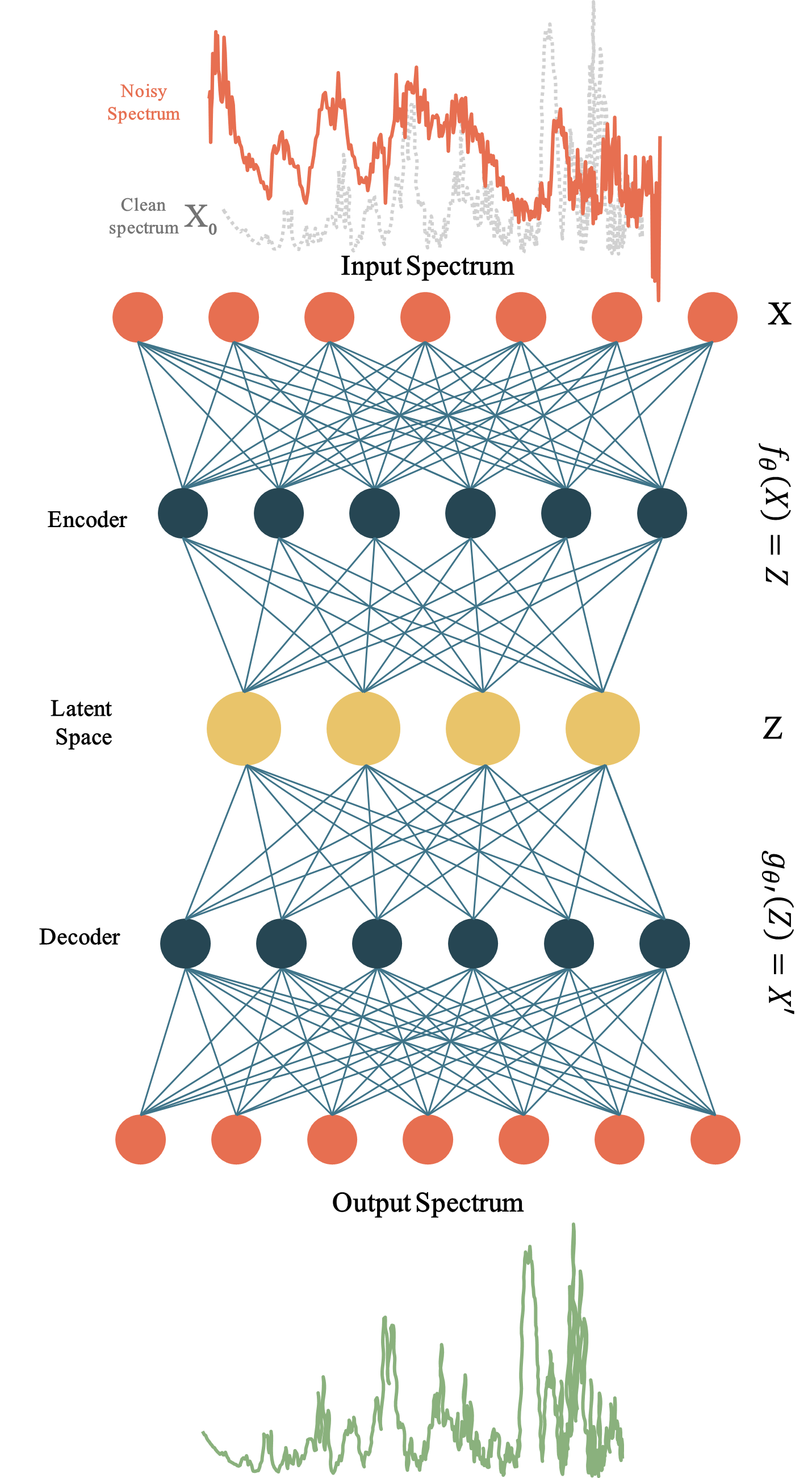}
    \caption{Architecture of a standard DAE. The input signal, $X$, is the observed transit spectrum (solid line), obtained by applying stellar contamination to the original uncontaminated spectrum, $X_0$ (dashed line). 
    The spectrum is passed through an encoder with multiple dense hidden layers that progressively compress the information into a compact latent representation, $Z$. The decoder then uses $Z$ to reconstruct a clean spectrum, $X'$. During training, the autoencoder learns to isolate and remove stellar contamination.}
    \label{fig:autoencoder_general}
\end{figure}

In general, besides the input and output layers, an autoencoder consists of two core components (see \autoref{fig:autoencoder_general} for a schematic representation): an encoder network, $f_\theta(X)$, characterized by a set of parameters, $\theta$, which reduces the input spectrum, $X$, to a compact latent representation, $Z$; and a decoder network, $g_{\theta'}(Z)$, that produces an output spectrum, $X'$. During training, the input and output spectra can be compared using various criteria. In denoising, the comparison is made between the output spectrum and the signal before noise and contamination (gray curve below the input spectrum in \autoref{fig:autoencoder_general}).

From a theoretical standpoint, autoencoders aim to approximate the identity function: given an input spectrum, the network should produce an output signal that is nearly identical to the input. This is achieved by incorporating constraints like bottleneck layers (e.g., reduction in the dimension of the input signal) or regularization terms (e.g., additional terms in the loss function that constrain properties of the latent space) to prevent trivial or overly simplistic solutions \citaeg{see e.g.,}{berahmandAutoencodersTheirApplications2024}. 

From a spectral analysis perspective, the encoder transforms a spectrum, potentially containing hundreds of channels (even hundreds of thousands of channels in some applications), into a reduced set of latent features that capture the essential information of the original spectrum \citep{scourfieldDenoisingGalaxyOptical2023a, melchiorAutoencodingGalaxySpectra2023a}. This latent representation often encapsulates the intrinsic dimensionality, i.e., the signal degrees of freedom (e.g., composition, mixing ratios, etc.) and underlying physical properties of the source \citaeg{see e.g.,}{kjaersgaardTAUNeuralNetwork2023}. Subsequently, the decoder reconstructs the signal from this compressed representation. Training an autoencoder involves minimizing a reconstruction loss function, such as a mean squared error (MSE), thereby simultaneously optimizing the parameters $\theta$ and $\theta'$ (e.g., weights and biases of neurons) of both the encoder and decoder.

Within the broader family of AE networks, DAEs are specifically designed to handle noisy or corrupted inputs. During training, these networks are intentionally exposed to corrupted data and are asked to reconstruct the original, clean version.
\footnote{It is important to stress that DAEs require clean target spectra during training; for transmission spectroscopy, this generally means training on synthetic data.}
The theoretical rationale behind DAEs is that forcing the network to recover clean data from noisy versions helps it learn robust, invariant features rather than superficial or irrelevant variations present in raw data. This method effectively isolates and removes noise, preserving and often highlighting essential features \citep{morvanDontPayAttention2022a, berahmandAutoencodersTheirApplications2024}. A key characteristic of DAEs is their implicit ability to learn the conditional distribution of clean data given noisy observations, making them particularly valuable in astronomy, where observational data frequently suffer from various sources of instrumental, atmospheric, or photon noise \citep{ghellerConvolutionalDeepDenoising2021}.

The result of applying the DAE is notable, as illustrated in \autoref{fig:spectrum_cleaning} in the main text. Beyond the underlying simplifications in the numerical experiment, the capability of a general DAE to reconstruct an uncontaminated spectrum without explicitly modeling contamination and noise motivates us to explore the power of DAEs in more general and realistic settings, such as those we study in this work. 

Nevertheless, the flexibility and unsupervised nature of autoencoders, together with recent advances in neural architectures—such as attention mechanisms and convolutional layers \citep{morvanDontPayAttention2022a, sedaghatStellarKaraokeDeep2023a}—make autoencoders powerful tools for addressing the increasing complexity and scale of astronomical datasets. Their capability to extract compact, physically meaningful representations from noisy observations shows great promise for upcoming large-scale surveys and next-generation astronomical instrumentation.

\section{Stellar contamination formalism}
\label{app:stellar_contamination}

In \autoref{sec:stellar_contamination} we presented the formula underpinning the stellar contamination model employed to construct our training set. In this section, we reproduce the derivation of the stellar contamination expression as a useful guide for directing future refinements of the model. 

We begin by calculating the observed transit depth, $D_\mathrm{obs}(\lambda)$, defined in terms of the flux blocked by the planet along the transit chord,
$\Delta F_\mathrm{chord}$, and the total emergent flux from the stellar photosphere, $F_\mathrm{out}$, given by

\begin{equation}
D_\mathrm{obs}(\lambda) \equiv \left(\frac{R_p}{R_\star}\right)^2_\mathrm{obs} = \frac{\Delta F_\mathrm{chord}(\lambda)}{F_\mathrm{out}(\lambda)}.
\label{eq:Dobs}
\end{equation}

We assume that the host star has an area fraction, $f_\mathrm{spot}$, covered by cooler stellar spots. For simplicity, we model the emergent specific flux from the spot component as a stellar spectrum evaluated at an effective temperature of $T_\mathrm{spot}<T_\mathrm{phot}$, where $T_\mathrm{phot}$ is the quiet photospheric temperature. In practice, these component spectra can be obtained by interpolating a stellar-atmosphere model grid (e.g., PHOENIX) at the corresponding effective temperatures. On the other hand, the stellar surface may also include a fraction, $f_\mathrm{fac}$, covered by bright faculae, which we model analogously as a stellar spectrum evaluated at an effective temperature of $T_\mathrm{fac}>T_\mathrm{phot}$. Taking into account these heterogeneities, the integrated flux of the star can be written as

\begin{equation}
\begin{aligned}
F_\mathrm{out}(\lambda) = & (1-f_\mathrm{spot}-f_\mathrm{fac}) F_\mathrm{phot}(\lambda,T_\mathrm{phot}) + \\
& + f_\mathrm{spot} F_\mathrm{spot}(\lambda,T_\mathrm{spot}) + f_\mathrm{fac} F_\mathrm{fac}(\lambda,T_\mathrm{fac}).
\end{aligned}
\label{eq:Fout}
\end{equation}

On the other hand, and similarly, the mean specific flux behind the chord is

\begin{equation}
\begin{aligned}
F_\mathrm{chord}(\lambda) = & (1-c_\mathrm{spot}-c_\mathrm{fac}) F_\mathrm{phot}(\lambda,T_\mathrm{phot}) + \\
& + c_\mathrm{spot} F_\mathrm{spot}(\lambda,T_\mathrm{spot}) + c_\mathrm{fac} F_\mathrm{fac}(\lambda,T_\mathrm{fac}),
\end{aligned}
\label{eq:Fchord}
\end{equation}
where $c_\mathrm{spot}$ and $c_\mathrm{fac}$ are the fractions of the surface covered by heterogeneities across the transit chord.

The change in flux produced by a planet with a projected radius, $R_p(\lambda)$, is

\begin{equation}
\Delta F_\mathrm{chord}(\lambda) = \left(\frac{R_p(\lambda)}{R_\star}\right)^2 F_\mathrm{chord}(\lambda).
\label{eq:DeltaFchord}
\end{equation}

Plugging Equations \ref{eq:DeltaFchord}, \ref{eq:Fchord}, and \ref{eq:Fout} into  \autoref{eq:Dobs}, we get

\begin{equation}
\left(\frac{R_p}{R_\star}\right)^2_\mathrm{obs} = \left(\frac{R_p(\lambda)}{R_\star}\right)^2 \epsilon_\lambda({\cal C}),
\end{equation}
where ${\cal C}=\{f_\mathrm{spot}, c_\mathrm{spot}, T_\mathrm{spot}, f_\mathrm{fac}, c_\mathrm{fac}, T_\mathrm{fac}\}$ is the complete set of  free parameters that characterize the TLS, and the function $\epsilon_\lambda$ is given by

\begin{equation}
\epsilon_{\lambda}({\cal C}) = \frac{(1 - c_{\mathrm{spot}} - c_{\mathrm{fac}}) F_\mathrm{phot}(\lambda) + c_{\mathrm{spot}} F_\mathrm{spot}(\lambda) + c_{\mathrm{fac}} F_\mathrm{fac}(\lambda)}{(1 - f_{\mathrm{spot}} - f_{\mathrm{fac}}) F_\mathrm{phot}(\lambda) + f_{\mathrm{spot}} F_\mathrm{spot}(\lambda) + f_{\mathrm{fac}} F_\mathrm{fac}(\lambda)}.
\end{equation}

\begin{figure*}
    \centering
    \includegraphics[width=0.65\linewidth]{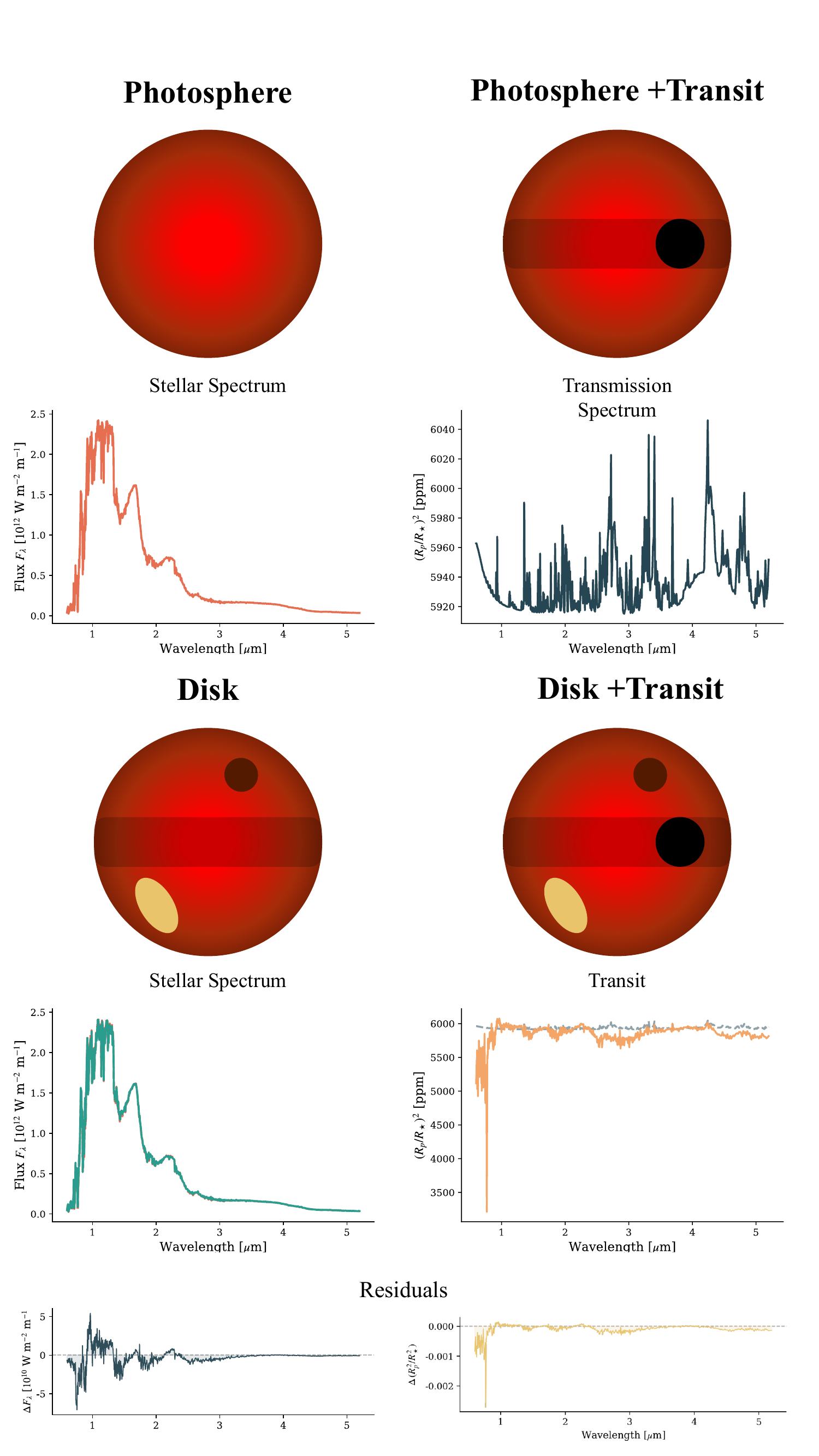}
    \caption{Schematic representation of the effect of stellar contamination (TLS) with actual examples of the effect of TLS in simulated signals. In the left column, we show the stellar spectrum, both when the photosphere is clean (no heterogeneities, stellar spots, or faculae) and when the star exhibits heterogeneities (second and third rows). The bottom row shows the residuals between the clean and contaminated spectra in each column. The residuals have been amplified relative to the spectra to highlight regions where the effects are more pronounced. In the right column, we show the corresponding transmission spectra: in the upper half, the homogeneous photosphere case, and in the bottom rows, the contaminated case. In all cases, we have illustrated the simpler case when the chord does not include any heterogeneity, $c_\mathrm{spot}=c_\mathrm{fac}=0$ (see text).}
    \label{fig:stellar_contamination}
\end{figure*}

\section{Uncontaminated model spectra}
\label{app:uncontaminated_spectra}

In \autoref{sec:G-DAE} and \autoref{sec:other_planets}, we presented the two planetary prototypes used throughout this work to train and evaluate our algorithms. While the broad range of possible compositions makes it impossible to define a single representative spectrum for the model planet, in this section we show two example spectra that help illustrate the main spectral characteristics present in our training set.

\begin{figure*}
    \centering    \includegraphics[width=0.8\linewidth]{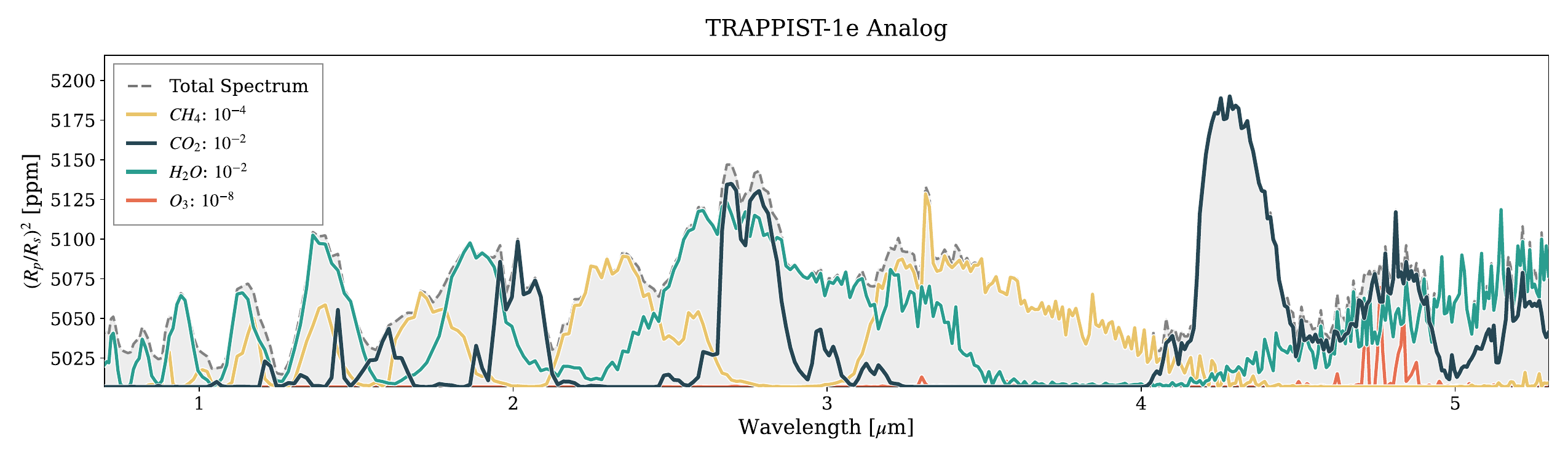}
    \caption{Uncontaminated transmission spectrum of a \exo{TRAPPIST-1}{}{e} analog with an arbitrary, non-equilibrium composition. Each gas contribution is shown as a thick continuous line; the fill gases (CO$_2$ and N$_2$) appear as thinner background lines. The combined spectrum is bounded by the dashed line and the shaded region shown in the background.}
    \label{fig:trappist1e_analogue_spectrum}
\end{figure*}

First, in \autoref{fig:trappist1e_analogue_spectrum}, we present a model spectrum for a representative \exo{TRAPPIST-1}{}{e} analog. In this example, we adopt atmospheric mixing ratios comparable to those inferred for Earth's surface during the Proterozoic eon \citep{kalteneggerHighresolutionTransmissionSpectra2020}.

\begin{figure*}
    \centering    \includegraphics[width=0.8\linewidth]{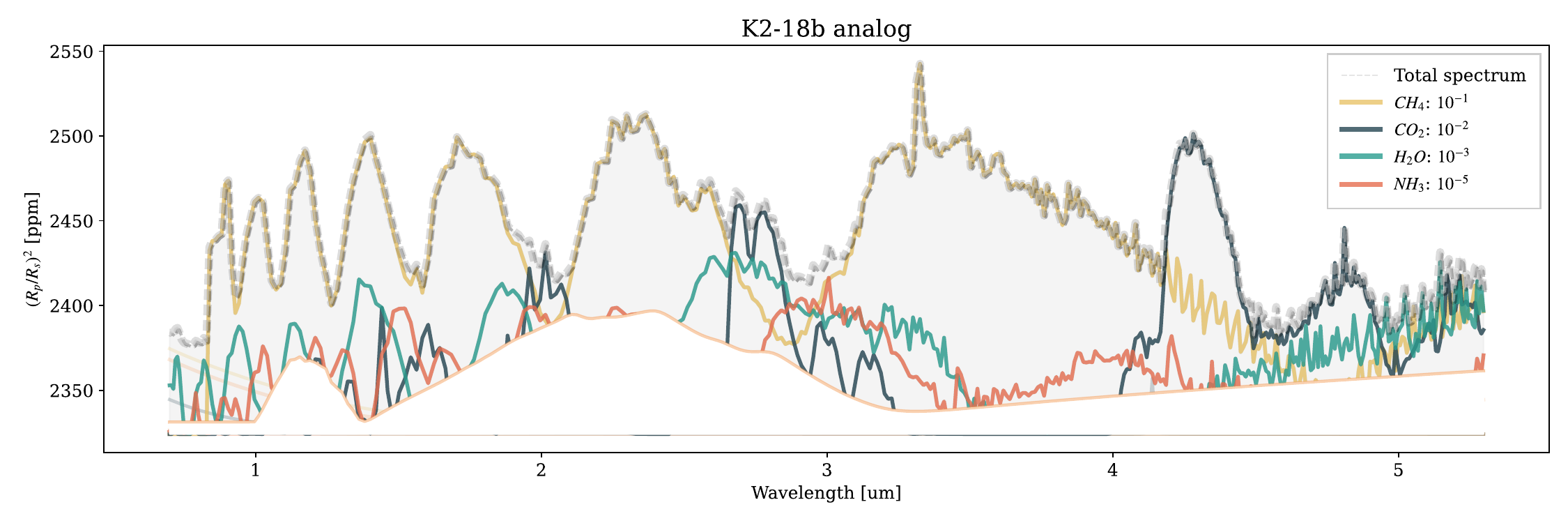}
    \caption{Uncontaminated transmission spectrum of a \exo{K2}{-18}{b} analog with an arbitrary (non-equilibrium) atmospheric composition. The contribution of each gas has been plotted as a continuous thick line, while the signal of the fill gases (H$_2$ and He, and their corresponding CIA) is shown as the shaded white curve below. The combined spectrum is limited by the dashed line and the shaded area in the background.}
    \label{fig:K2-18b_analogue_spectrum}
\end{figure*}

Secondly, in \autoref{fig:K2-18b_analogue_spectrum}, we present the synthetic, contamination-free spectrum of a mock sub-Neptune whose atmospheric composition falls within the parameter ranges adopted for training our algorithms. In this case, the specific mixing ratios were chosen to highlight the most salient spectral signatures of the chemical species and do not necessarily reflect the equilibrium abundances of the molecules. We can distinguish in the mock spectrum a wide CH$_4$ band around 3.3 $\mu$m, which dominates the spectrum in that region. Also noticeable are the bands of CO$_2$ around 2, 2.8, 4.3, and 4.8 $\mu$m. At least one band of NH$_3$ may be salient in the region of 3 $\mu$m, while water spectral signatures dominate the redder wavelength regions.

It should be emphasized that none of these spectra can be regarded as fully representative of the entire training set, which spans a broad range of compositions designed to supply the algorithm with sufficient information to reconstruct any test spectrum. \autoref{tab:Gen_DAE_dataset} summarizes the dataset composition across S/N levels and atmospheric configurations. Values are expressed in thousands and account for all noisy realizations and stellar contamination scenarios. The complete training set comprised $3{,}605{,}592$ spectra.

\begin{table}[h]
\centering
\caption{Distribution of training spectra for the G-DAE.}\label{tab:Gen_DAE_dataset}
\resizebox{\linewidth}{!}{%
\begin{tabular}{lccccccccc}
\hline
S/N & 000 & 100 & 010 & 001 & 110 & 101 & 011 & 111 & Airless \\ 
\hline
1  & 114   & 91.20 & 91.20 & 91.20 & 72.96 & 72.96 & 72.96 & 116.736 & 38 \\
3  & 114   & 91.20 & 91.20 & 91.20 & 72.96 & 72.96 & 72.96 & 116.736 & 38 \\
6  & 85.50 & 68.40 & 68.40 & 68.40 & 54.72 & 54.72 & 54.72 & 87.552  & 28.50 \\
10 & 57    & 45.60 & 45.60 & 45.60 & 36.48 & 36.48 & 36.48 & 58.368  & 19 \\
0  & 171   & 136.80 & 136.80 & 136.80 & 109.44 & 109.44 & 109.44 & 145.92 & 76 \\
\hline
\end{tabular}
}
\tablefoot{Each row corresponds to a different training-set S/N, with S/N = 0 indicating noise-free spectra. The columns labeled by three-digit binary codes indicate the presence (1) or absence (0) of CH$_4$, O$_3$, and H$_2$O. The ``Airless'' column contains flat, absorption-free spectra. Cell values, expressed in thousands, give the total number of spectra for each atmospheric configuration and noise level.}
\end{table}

\section{An N+1 test with DMS}
\label{sec:nplus1_dms}

In practical observations, we must anticipate unforeseen atmospheric compositions; for example, employing a G-DAE trained on spectra with at most N molecules when the observed spectrum actually includes an additional, previously unseen molecule. We refer to this as the $N+1$ scenario. To investigate this scenario, we applied the sub-Neptune G-DAE—originally trained only on CO$_2$, CH$_4$, H$_2$O, and NH$_3$—to synthetic spectra that included a range of DMS abundances. This molecule is a highly debated potential biosignature on \exo{K2}{-18}{b} \citep{madhusudhanNewConstraintsDMS2025}, making it an ideal test case. 

\begin{figure*}[ht!]
    \centering
    \includegraphics[width=0.6\linewidth]{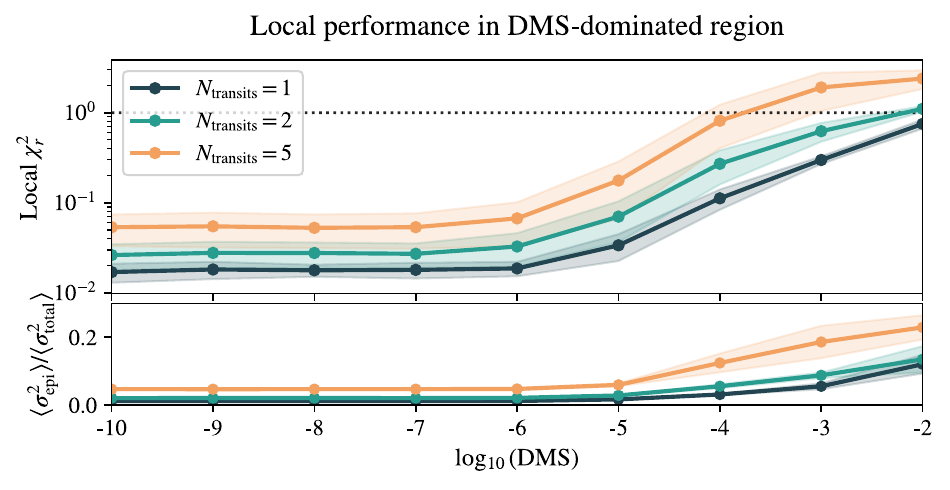}
    \caption{G-DAE performance on mock spectra with an unexpected molecule (DMS), evaluated in the DMS-dominated region ($\lambda \geq 4\,\mu\mathrm{m}$). \textbf{Top:} Local reduced chi-square ($\chi_r^2$) for one, two, and five transits. The values scale with DMS abundance and show that the reconstruction discrepancy is entirely due to the unmodeled gas. \textbf{Bottom:} Fractional epistemic contribution to the total predictive variance. Low values indicate high network confidence, whereas increased values indicate uncertainty when encountering unfamiliar spectral features. Solid lines show means; shaded bands show the full range across stellar contamination scenarios.}
    \label{fig:dms_nplus1}
\end{figure*}

Retaining our previous stellar contamination and \textsc{PandExo} noise models, we focused our analysis on the infrared region where DMS absorbs most strongly ($\lambda \geq 4\,\mu$m). To quantify the network's response, we measured both the reconstruction error (local reduced $\chi_r^2$) and the network's uncertainty about its prediction (epistemic uncertainty estimated via Monte Carlo Dropout). The corresponding results are presented in \autoref{fig:dms_nplus1}.

Because the G-DAE is constrained by the chemical library it learned during training, its reconstructions inevitably struggle to match the true $N+1$ spectra. However, the way the network fails is highly informative. When the DMS signal is weak—due to either trace abundances or noise-dominated observations (e.g., a single transit)—the network essentially masks the feature, maintaining a good fit ($\chi_r^2 < 1$) with high confidence. But as the DMS abundance and the S/N increase, the unexpected signature becomes prominent. Instead of confidently forcing the spectrum to look like a known gas, the G-DAE responds by proportionally inflating its epistemic uncertainty right in the mismatched wavelength band.

These results highlight the true value of the G-DAE as a decontamination preprocessor rather than a replacement for full retrievals. While the network cannot explicitly name the missing molecule, its inflated error bars act as an automated diagnostic warning for unexpected spectral features. By propagating these enlarged uncertainties into downstream retrievals, we naturally weaken the statistical weight of the anomalous band. This prevents the retrieval algorithm from confidently locking onto an incorrect chemical composition, ensuring that multiple atmospheric models remain statistically viable for the discovery of new physics.

\end{appendix}

\FloatBarrier
\clearpage
\end{document}